\documentclass[resetfootnote, trackchanges, twocolumn, twocolappendix]{aastex701}
\usepackage{amsmath}
\usepackage{bm}
\usepackage{booktabs} 
\usepackage{CJK}
\usepackage{comment}
\usepackage{txfonts} 

\usepackage{placeins}
\usepackage{wasysym}    
\usepackage{xspace}
\usepackage{xcolor}
\usepackage{soul}
\usepackage{natbib}
\usepackage{adjustbox}
\usepackage[version=4]{mhchem}

\usepackage{graphicx}
\usepackage{subcaption}
\usepackage[normalem]{ulem}

\shorttitle{\textsc{Arepo-CD}}
\shortauthors{C. Kong et al.}

\begin{document}
\begin{CJK*}{UTF8}{gbsn}

\title{\textsc{Arepo-CD}: Coupling Collisional $N$-Body Dynamics with Moving-Mesh Magnetohydrodynamics}

\author[0009-0003-9099-5805,gname=Chuizheng,sname=Kong]{Chuizheng Kong (孔垂政)}
\affiliation{Department of Astronomy, Tsinghua University, Beijing, 100084, People's Republic of China}
\email{kcz24@mails.tsinghua.edu.cn}

\author[0000-0002-1253-2763,gname=Hui,sname=Li]{Hui Li (李辉)}
\affiliation{Department of Astronomy, Tsinghua University, Beijing, 100084, People's Republic of China}
\email[show]{hliastro@tsinghua.edu.cn}

\author[0000-0001-8713-0366,gname=Long,sname=Wang]{Long Wang (王龙)}
\affiliation{School of Physics and Astronomy, Sun Yat-sen University, Daxue Road, Zhuhai, 519082, China}
\affiliation{CSST Science Center for the Guangdong-Hong Kong-Macau Greater Bay Area, Zhuhai, 519082, China}
\email{wanglong8@mail.sysu.edu.cn}

\author[0000-0002-7478-6427,gname=Yunwei,sname=Deng]{Yunwei Deng (邓云未)}
\affiliation{Department of Astronomy, Tsinghua University, Beijing, 100084, People's Republic of China}
\email{dengyw22@mails.tsinghua.edu.cn}

\author[0000-0002-6906-2379,gname=Elizabeth,sname=Moreno-Hilario]{Elizabeth Moreno-Hilario}
\affiliation{Department of Astronomy, Tsinghua University, Beijing, 100084, People's Republic of China}
\email{elimorh@tsinghua.edu.cn}


\begin{abstract}

Star formation is a hierarchical, clustered process in turbulent giant molecular clouds (GMCs), where close stellar encounters can strongly affect cluster evolution and feedback. However, gravitational softening in many hydrodynamic simulations of GMCs limits the treatment of these encounters. We present \textsc{Arepo-CD} (Collisional Dynamics), a framework that couples star-by-star collisional $N$-body dynamics to the moving-mesh magnetohydrodynamics (MHD) code \textsc{Arepo}. Star-star interactions use unsoftened gravity, while interactions involving gas or collisionless components remain softened. Compact stellar subsystems are identified on the fly and evolved with a fourth-order forward symplectic integrator and slow-down algorithmic regularization coupled to the native \textsc{Arepo} timestep hierarchy. Benchmark tests against the collisional $N$-body code \textsc{PeTar} show good agreement in cluster core collapse and subsequent expansion, radial density profiles, and stellar velocity distributions. Coupled to the \textsc{Rigel} star formation and feedback model, \textsc{Arepo-CD} follows cluster formation and early evolution in radiation MHD simulations of a $10^5~M_\odot$ turbulent molecular cloud. In the fiducial GMC comparison, the collisional and softened-gravity runs yield similar star formation histories and integrated star formation efficiencies. Nevertheless, close gravitational encounters during the embedded phase generate a substantial runaway population, with a velocity tail extending beyond $100~{\rm km/s}$. Ejected massive stars redistribute radiative feedback into the surrounding lower-density medium and produce a more extended distribution of heated gas. \textsc{Arepo-CD} provides a practical and computationally efficient framework for studying collisional stellar dynamics and its coupling to gas evolution and feedback during star cluster formation in realistic galactic environments.

\end{abstract}
\keywords{\uat{Star formation}{1569} --- \uat{Stellar dynamics}{1596} --- \uat{Young star clusters}{1833} --- \uat{Molecular clouds}{1072} --- \uat{Magnetohydrodynamical simulations}{1966} --- \uat{N-body simulations}{1083}}

\section{Introduction} \label{sect:intro}

Star clusters form in the densest regions of giant molecular clouds (GMCs), where supersonic turbulence, self-gravity, and magnetic fields produce a highly structured interstellar medium \citep{Larson1981, Lada&Lada2003, McKee&Ostriker2007}. Cloud collapse generates filaments, hubs, and compact overdensities in which stars form in a strongly clustered and hierarchical fashion \citep{Longmore+2014SCFinMCs, Krumholz+2019}. Observations identify star-forming regions at different evolutionary stages, from deeply embedded star-forming regions to exposed young massive clusters. These observations show that the transition is rapid and dynamically complex \citep{Schneider+2010SCFinMCs, Adamo+2020, Schinnerer&Leroy2024SCFinMCs}. During this phase, stars continue to form and inject feedback as the natal cloud is dispersed \citep{Hennebelle&Grudic2024SCFinMCs, Beuther+2025SCFinMCs}. Cluster formation therefore depends on the complex interplay among gas dynamics, star formation, feedback, and stellar encounters over a wide range of scales. The coupled, multiscale nature of these processes makes numerical simulations essential for a self-consistent theoretical description of star cluster formation.

On the one hand, hydrodynamic simulations provide a natural framework for studying this process because the gaseous environment strongly affects the properties of forming stars \citep[e.g.][]{Dale+2011SCFinMCs, Dale+2012SCFinMCs, Kim+2018SCFinMCs, Howard+2018SCFinMCs, Li+2019, Fukushima+2020SCFinMCs, Grudic+2021SCFinMCs}. Current simulations can follow the formation and evolution of individual stars in molecular clouds and star-forming regions, from the initial cloud collapse to the dispersal of the natal gas \citep[e.g.][]{Bate2012SCFinMCs, Bate2014SCFinMCs, Bate2019SCFinMCs, He+2019SCFinMCs, Guszejnov+2021STARFORGE, Grudic+2021STARFORGE, Grudic+2022STARFORGE, Farias+2024STARFORGE}. However, most such simulations use softened stellar gravity and have limited spatial resolution, which restrict their treatment of collisional stellar dynamics. This limitation artificially suppresses close encounters, binary formation and hardening, few-body scattering, dynamical ejections, and rapid mass segregation. This can alter both the internal evolution of a cluster and the locations at which massive stars inject feedback.

On the other hand, collisional $N$-body simulations provide a complementary approach to modeling stellar encounters \citep[e.g.][]{Wang+2015NBody6++GPU, Wang2020PeTar, Rantala+2023BIFROST}. Here, ``collisional'' refers to stellar systems in which few-body gravitational encounters are dynamically important. With high force accuracy and specialized integration methods, $N$-body calculations can follow relaxation, core contraction, binary-mediated energy exchange, dynamical ejections, and long-term structural evolution \citep{Baumgardt+2002NBody}. They have also clarified how initially substructured stellar distributions evolve and how strong encounters reshape young clusters \citep{Goodwin&Whitworth2004SC}. Their main limitation during the embedded phase is that they do not follow star formation self-consistently in an evolving gaseous environment. Conventional models begin with prescribed stellar initial conditions (ICs). They either neglect the natal gas or represent it with a static or time-dependent background potential \citep{Farias+2019runaway&DES, Schoettler+2019runaway&DES&walkaway, Dinnbier&Kroupa2020, Dinnbier+2022}. These models therefore have to assume, rather than predict, the timing and geometry of gas dispersal.

Several hybrid ``hydro + $N$-body'' approaches combine hydrodynamic solvers with accurate collisional dynamics \citep[e.g.][]{Wall+2019FLASH+ph4, Polak+2024aTorch, Fujii+2021bASURA+BRIDGE, Fujii+2024aASURA+BRIDGE, Jo+2024Enzo-Abyss, Lahen+2025aSPHGAL+KETJU, Bernard+2025PhantomN, Chung+2026Enzo-Abyss, Jo+2026Enzo-Abyss}. These approaches either couple specialized few-body solvers through external interfaces or embed stellar-dynamics algorithms directly within hydrodynamic codes. Interface-based methods are flexible and can combine mature solvers, but require frequent communication and synchronization between codes \citep[e.g.][]{Wall+2019FLASH+ph4, Polak+2024aTorch, Fujii+2021bASURA+BRIDGE, Fujii+2024aASURA+BRIDGE}. Alternatively, embedded methods share the gravity solver, timestep hierarchy, and parallel infrastructure of the hydrodynamic code, although existing implementations differs in their treatments on collisional gravity or coupled baryonic physics \citep{Lahen+2025aSPHGAL+KETJU, Bernard+2025PhantomN, Jo+2024Enzo-Abyss, Chung+2026Enzo-Abyss, Jo+2026Enzo-Abyss}.

The main numerical challenge in such calculations is the wide range of spatial and temporal scales. Molecular-cloud evolution proceeds over several Myr, whereas tight binaries and close encounters can have characteristic timescales as short as days. Removing gravitational softening restores the force accuracy required for these systems, but produces an extremely deep timestep hierarchy in a conventional integrator. An efficient method must therefore separate compact few-body dynamics from the slower evolution of the surrounding cloud and apply high-accuracy integration only where it is needed.

Motivated by this requirement, we present \textsc{Arepo-CD}, a framework that couples star-by-star collisional $N$-body dynamics to the moving-mesh magnetohydrodynamics (MHD) code \textsc{Arepo} \citep{Springel2010Arepo, Pakmor+2011Arepo,Pakmor&Springel2013Arepo, Pakmor+2016Arepo,Weinberger+2020Arepo}. The framework treats star--star gravity without softening. Interactions involving gas or collisionless components remain softened. \textsc{Arepo-CD} combines hierarchical time-stepping with on-the-fly identification of compact stellar groups. It evolves their internal dynamics with a fourth-order forward symplectic integrator and slow-down algorithmic regularization \citep{Oshino2011_P3T, Wang+2020SDAR, Rantala+2021FROST,Rantala+2023BIFROST}. This specialized treatment is applied to dynamically demanding subsystems. The rest of the calculation remains within the native \textsc{Arepo} framework and is compatible with its other physics modules. We couple \textsc{Arepo-CD} to the \textsc{Rigel} radiation MHD model for star formation and feedback from individual stars \citep{Deng+2024RIGEL, Deng+2025RIGEL, Deng+2026aRIGEL, Deng+2026bRIGEL}. This coupling allows embedded clusters to form during GMC collapse and evolve under the simultaneous influence of gas dynamics, feedback, and stellar encounters, while resolving binary formation, strong scattering, and dynamical ejections at an efficient computational cost suitable for hydrodynamic applications.

In this paper, we describe the numerical framework of \textsc{Arepo-CD} and present its first astrophysical applications. In Sect.~\ref{sect:method}, we describe the hybrid gravity treatment, integration scheme, timestep criteria, group-search algorithm, and coupled star formation and feedback model. In Sect.~\ref{sect:test}, we validate the method against collisional $N$-body benchmarks. In Sect.~\ref{sect:hydro}, we present the first GMC applications. In Sect.~\ref{sect:discuss}, we discuss related frameworks, numerical performance, limitations, and future directions. In Sect.~\ref{sect:summary}, we summarize our conclusions.

\section{Numerical Methods} \label{sect:method}

In this section, we describe the main numerical methods implemented in \textsc{Arepo-CD}, which incorporates collisional $N$-body dynamics into the moving-mesh MHD code \textsc{Arepo}. \textsc{Arepo} provides an efficient tree-gravity solver, moving-mesh hydrodynamics, and a broad set of physics modules suitable for simulations of star clusters in gaseous environments \citep{Li+2019, Deng+2025RIGEL, Smith2026SFmodel}.\footnote{The numerical methods developed here could, in principle, be implemented in other star-by-star hydrodynamic frameworks using direct summation, octree gravity, or particle-mesh gravity.} However, the standard \textsc{Arepo} gravity solver is designed for collisionless $N$-body dynamics. Close binaries, hyperbolic encounters, triples, and higher-order multiples can impose extremely short dynamical timescales. Direct integration of these systems within the timestep hierarchy can therefore become inefficient or prohibitively expensive \citep{Springel2005GADGET-2, Springel2010Arepo, Weinberger+2020Arepo, Springel2021GADGET-4}.

\textsc{Arepo-CD} addresses this problem by identifying compact stellar subsystems on-the-fly and separating their internal few-body evolution from the large-scale hydrodynamics. We first outline the overall framework and hybrid gravity strategy in Sect.~\ref{subsect:overview}, after which the remainder of this section presents the main numerical ingredients of \textsc{Arepo-CD}.

\begin{figure}[htbp!] \centering
\includegraphics[width=\linewidth]{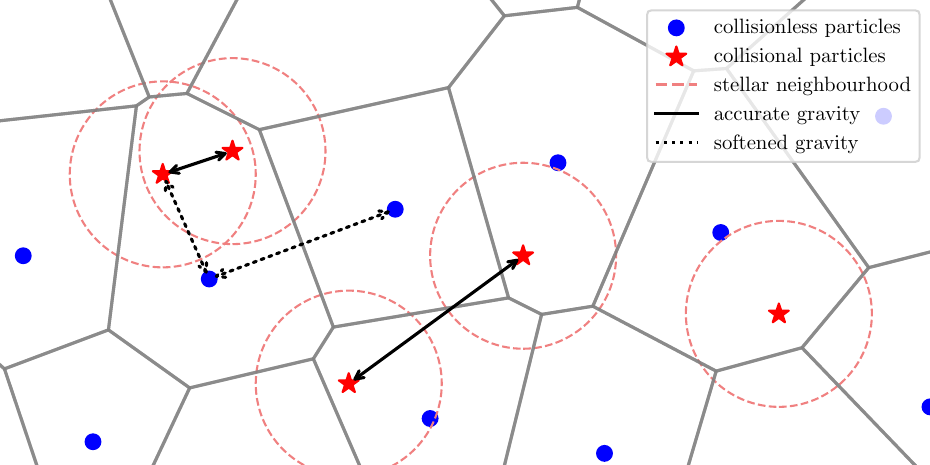}
\caption{Schematic of the hybrid gravity treatment in \textsc{Arepo-CD}. Blue symbols denote collisionless components (gas, dark matter, etc.), and red symbols denote collisional stars or stellar remnants. Arrow labels specify the gravitational model used for each interaction type: softened gravity for any interaction involving a collisionless particle, and Newtonian gravity for interactions between stellar particles. Dashed red circles mark the neighbor regions of collisional particles, within which the tree is force-opened to resolve close stellar encounters (see Sect.~\ref{subsubsect:pair}).}
\label{fig:schematic_diagram}
\end{figure}

\subsection{Overview of the Framework} \label{subsect:overview}

We begin by summarizing the modified hybrid gravity treatment, which Fig.~\ref{fig:schematic_diagram} illustrates. Particles are divided into collisionless and collisional components. The collisional component, including individual stars and stellar remnants, interacts internally through unsoftened Newtonian gravity, while all interactions involving the collisionless component, including gas, dark matter, and other species, use softened gravity. To implement this hybrid treatment, we modify the Barnes--Hut octree \citep{Barnes&Hut1986octree} in \textsc{Arepo} so that each tree node stores the collisional and collisionless mass components separately. In addition, when the pairwise dynamical timescale of two stellar particles becomes sufficiently short, they are assigned to the same collisional ``group'' (see Sects.~\ref{subsubsect:pair}, \ref{subsubsect:P3T}, and \ref{subsubsect:search} for the timescale criterion, group definition, and group-search procedure, respectively). The internal dynamics of each group are then evolved with a specialized few-body solver (Eq.~\ref{eq:FSI+SDAR}). During group integration, the gravitational influence of particles outside the group, including the remaining stellar particles and collisionless components, is approximated as a constant external background. This influence is accounted for separately through the long-range kicks (Sect.~\ref{subsubsect:P3T}), rather than during the group integration.

\begin{figure}[htbp!] \centering
\includegraphics[width=\linewidth]{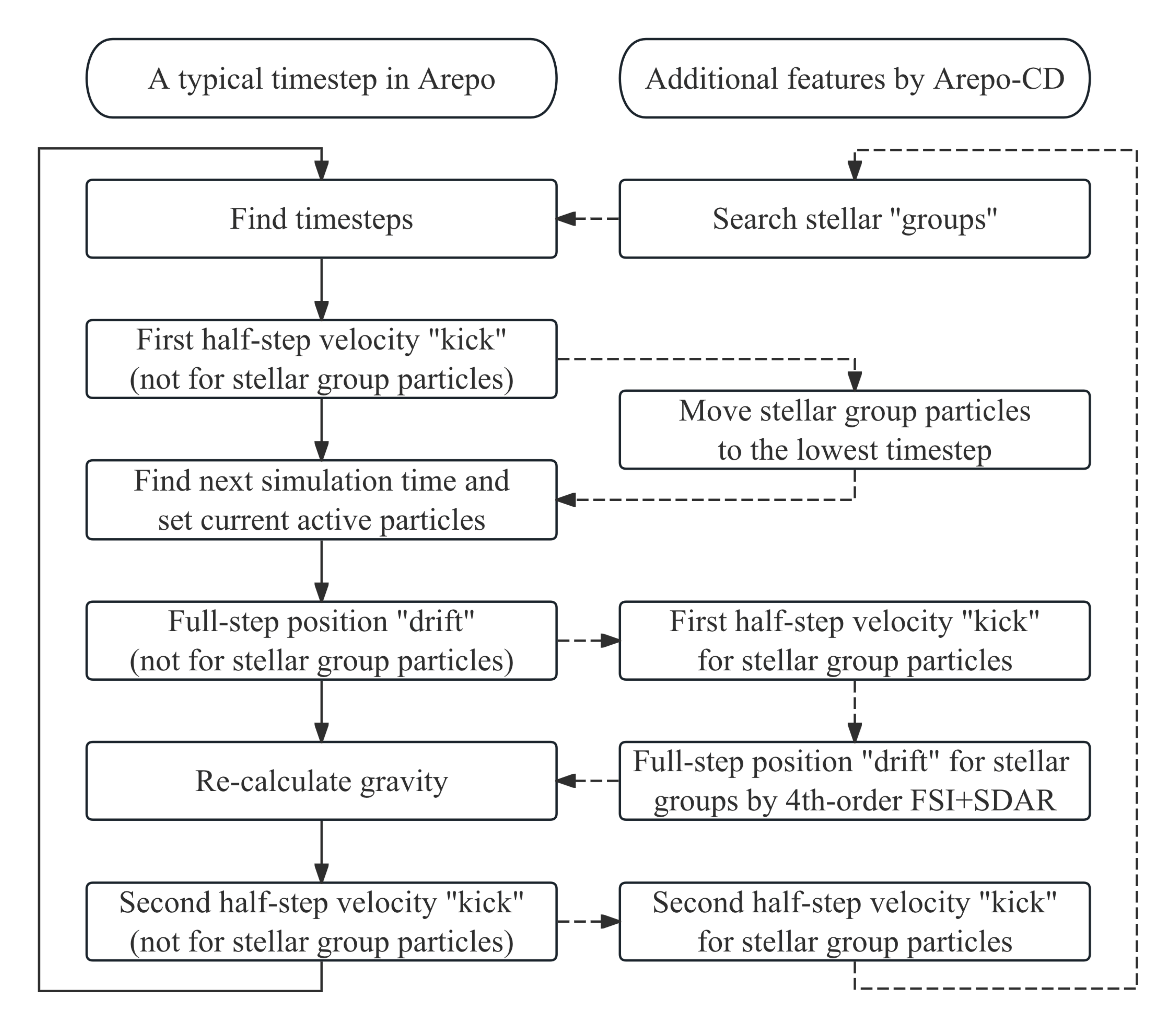}
\caption{Flow chart of the operations performed during a typical timestep. Solid lines show the standard \textsc{Arepo} workflow, while dashed lines indicate additional steps executed when the \textsc{Arepo-CD} module is enabled. The schematic is simplified and includes only the dynamical integration. The definition and identification of stellar ``groups'' appear in Sects.~\ref{subsubsect:P3T} and \ref{subsect:group}, respectively. The velocity ``kick'' and position ``drift'' correspond to the leapfrog steps reviewed in Sect.~\ref{subsubsect:leapfrog}. The ``4th-order FSI'' and ``SDAR'' are introduced separately in Sects.~\ref{subsubsect:4thFSI} and \ref{subsubsect:SDAR}, respectively.}
\label{fig:flow_chart}
\end{figure}

Fig.~\ref{fig:flow_chart} illustrates the \textsc{Arepo-CD} workflow for dynamical integration only. Below, we describe the additional steps indicated by dashed lines when the \textsc{Arepo-CD} module is enabled. At the end of each timestep, the force tree walk distinguishes unsoftened stellar interactions from softened interactions and also records stellar-neighbor information for subsequent group identification (Sect.~\ref{subsubsect:pair}). At the beginning of the next timestep, this information is used to identify compact stellar groups (Sect.~\ref{subsubsect:search}). We then apply the revised timestep criteria for collisional stellar particles in the groups (Sect.~\ref{subsect:timestep}). Particles in the identified groups are subsequently reassigned to the smallest timestep currently present in the timestep hierarchy to synchronize them with the other particles. The groups are then evolved in parallel using the hybrid integration scheme described in Sect.~\ref{subsect:integration}, which combines a fourth-order forward symplectic integrator and slow-down algorithmic regularization. This resolves internal group dynamics without imposing the shortest internal timesteps on the entire simulation. Finally, Sect.~\ref{subsect:RIGEL} describes the coupling between \textsc{Arepo-CD} and the \textsc{Rigel} star formation and feedback model used in the hydrodynamic applications.

\subsection{Time Integration} \label{subsect:integration}

Star-by-star cluster formation simulations span a wide range of spatial and temporal scales, from the global cluster evolution to close encounters and compact few-body systems such as binaries and triples. This range challenges the standard \textsc{Arepo} leapfrog integrator. For example, a stable tight binary with an orbital period below one day must be followed over several Myr. This requires a timestep hierarchy of order Myr/day $\sim 10^9$. Resolving such systems directly with the default global integration framework would be inefficient, because even an update involving only a few stellar particles can require repeated force-tree walks and thereby slow the entire simulation. Moreover, the default scheme is only second-order accurate, and accumulated truncation errors over many orbits can artificially disrupt hard binaries or cause spurious cluster expansion \citep{Aarseth2003}. Modern collisional stellar-dynamics codes therefore typically use fourth- or higher-order methods \citep{Wang+2015NBody6++GPU, Wang2020PeTar, Rantala+2023BIFROST}.

To address this problem, no single integration method can provide both high accuracy and high efficiency across the full dynamical range of collisional systems. We therefore combine complementary methods into a hybrid Hamiltonian-splitting scheme with three levels (Sect.~\ref{subsubsect:P3T}): (1) leapfrog, the native \textsc{Arepo} scheme, advances the large-scale evolution and handles interactions between collisional stars and collisionless components (Sect.~\ref{subsubsect:leapfrog}); (2) the forward symplectic integrator (FSI) accurately evolves the internal dynamics of compact stellar groups (Sect.~\ref{subsubsect:4thFSI}); and (3) slow-down algorithmic regularization (SDAR) treats the stiffest close interactions efficiently, particularly hard binaries and triples (Sect.~\ref{subsubsect:SDAR}). To improve efficiency, independent compact stellar groups are distributed across parallel tasks and assigned group timesteps longer than their shortest internal dynamical timescales (Sect.~\ref{subsubsect:parallel}).\footnote{For clarity, we denote particle masses by $m$, positions by $\bm{r}$, velocities by $\bm{v}$, and momenta by $\bm{p}$. The phase-space vector is defined as $\bm{w} = (\bm{r}, \bm{p})$. A subscript index (e.g., $i, j, 1, 2, 3, \dots$) identifies a specific particle, whereas the absence of a subscript refers to all particles within the $N$-body system. For example, $\bm{w} = (\bm{w}_1, \bm{w}_2, \dots, \bm{w}_N)$.}

\subsubsection{Leapfrog Integrator} \label{subsubsect:leapfrog}

We first briefly review the basic leapfrog integrator used in \textsc{Arepo}, written in operator-splitting form \citep{Springel2005GADGET-2, Springel2021GADGET-4, Mannerkoski2023_KETJU}, as a foundation for the operator formulations introduced below. For a Hamiltonian system, time evolution over a timestep $\Delta t$ can be approximated using the standard Hamiltonian splitting $H = T + V$, where $H$ is the Hamiltonian, and $T$ and $V$ are the kinetic and potential energies, respectively \citep{Yoshida1990}. The corresponding canonical maps are the drift ($D$) and kick ($K$) operators, respectively. The second-order, time-symmetric Strang splitting gives the standard kick-drift-kick (KDK) leapfrog update,
\begin{align}
E (\Delta t) \approx K \left( \dfrac{\Delta t}{2} \right) D (\Delta t) K \left( \dfrac{\Delta t}{2} \right) ,
\label{eq:KDK}
\end{align}
where $K (\Delta t / 2)$ is a half-step ``kick'' that updates momenta, or equivalently velocities, under the potential force. The full-step ``drift'' $D (\Delta t)$ updates positions under the kinetic term.

This KDK leapfrog is time-reversible and symplectic, making it efficient and robust for long-term gravitational integrations, with bounded energy oscillations rather than the secular drift often seen in non-symplectic schemes (e.g., Runge-Kutta; \citealt{Springel2005GADGET-2}). We note that a different Hamiltonian splitting leads to different operators, as discussed in Sect.~\ref{subsubsect:P3T}, where we introduce an alternative splitting scheme.

\subsubsection{Fourth-order Forward Symplectic Integrator} \label{subsubsect:4thFSI}

We adopt a fourth-order symplectic integrator with strictly positive substeps to improve the integration of close encounters and compact few-body systems \citep{Yoshida1990, Chin1997FSI4th, Chin2007FSI4th, Xu&Wu2010FSI4th, Hernandez&Bertschinger2015NBody, Dehnen&HernandezFSI4th, Hernandez&Bertschinger2018NBody, Rantala+2021FROST, Hernandez&Dehnen2023NBody, Rantala+2023BIFROST}. Following \citet{Rantala+2021FROST, Rantala+2023BIFROST} and \citet{Bernard+2025PhantomN}, we write the forward symplectic integrator (FSI) as a five-stage map over one timestep $\Delta t$, consisting of a kick, drift, gradient-corrected kick, drift, and kick. Using the drift operator $D (\Delta t)$, the kick operator $K (\Delta t)$, and the gradient-corrected kick operator $\tilde{\mathcal{K}} (\Delta t)$, this cycle can be written as
\begin{align}
E_{\rm FSI} (\Delta t) = K \left( \dfrac{\Delta t}{6} \right) D \left( \dfrac{\Delta t}{2} \right) \tilde{\mathcal{K}} \left( \dfrac{2}{3} \Delta t \right) D \left( \dfrac{\Delta t}{2} \right) K \left( \dfrac{\Delta t}{6} \right) .
\label{eq:FSI}
\end{align}
The kick, drift, and gradient-corrected kick update the particles as follows:
\begin{align}
K \left( \dfrac{\Delta t}{6} \right): \quad \bm{v}_i & \rightarrow \bm{v}_i + \dfrac{1}{6} \bm{a}_i \Delta t, \label{eq:kick}\\
D \left( \dfrac{\Delta t}{2} \right): \quad \bm{r}_i & \rightarrow \bm{r}_i + \dfrac{1}{2} \bm{v}_i \Delta t, \label{eq:drift}\\
\tilde{\mathcal{K}} \left( \dfrac{2}{3} \Delta t \right): \quad \bm{v}_i & \rightarrow \bm{v}_i + \dfrac{2}{3} \bm{a}_i \Delta t \notag \\
& \quad + \dfrac{1}{36}\sum_{j=1,\,j\neq i}^{N}\dfrac{G m_j}{r_{ji}^3} \left( \bm{a}_{ji} - 3 \dfrac{\bm{a}_{ji} \cdot \bm{r}_{ji}}{r_{ji}^2} \bm{r}_{ji} \right) \Delta t^3 .
\label{eq:mod_kick}
\end{align}
Here $\bm{a}_i$ is the Newtonian gravitational acceleration of particle $i$ and $G$ is the gravitational constant. We define $\bm{r}_{ji} \equiv \bm{r}_j - \bm{r}_i$ with $r_{ji} \equiv |\bm{r}_{ji}|$, and $\bm{a}_{ji} \equiv \bm{a}_j - \bm{a}_i$.

The gradient-corrected kick in Eq.~\ref{eq:mod_kick} makes the FSI composition fourth-order accurate. Its $\mathcal{O} (\Delta t^3)$ correction cancels the leading error terms of the second-order scheme. Compared with high-order symplectic integrators in \cite{Yoshida1990}, the fourth-order FSI uses strictly positive sub-step lengths, which is advantageous for irreversible dynamics (e.g., with gravitational-wave dissipation) and can improve the efficiency of hierarchical recursive integration. The method also performs better for the Kepler problem than schemes with negative substeps \citep{Chin1997FSI4th}.

\subsubsection{Slow-Down Algorithmic Regularization} \label{subsubsect:SDAR}

Increasing the integration order does not remove all difficulties associated with close encounters and compact few-body systems, since highly eccentric binary stars and close hyperbolic encounters can still impose prohibitively small timesteps. We therefore include slow-down algorithmic regularization\footnote{GitHub repository: \url{https://github.com/lwang-astro/SDAR}} (SDAR; \citealt{Wang+2020SDAR}), a few-body integration method that combines two complementary methods. The slow-down (SD; \citealt{Mikkola&Aarseth1996SD}) technique reduces the computational cost for weakly perturbed hard binaries by artificially slowing their internal orbital phase while preserving the secular effects of external perturbations. Algorithmic regularization (AR; \citealt{Mikkola&Tanikawa1999AR, Preto&Tremaine1999AR}) is a time-transformed explicit symplectic scheme that improves robustness and conservation during close encounters.

For a perturbed binary, the SD technique rescales only the internal binary motion while leaving the center-of-mass (CoM) motion unchanged. In the laboratory frame, the equations of motion for component $i = 1, 2$ with companion $j \neq i$ are
\begin{align} \begin{cases}
\dfrac{\mathrm{d} \bm{r}_i}{\mathrm{d} t} = \dfrac{1}{\kappa} ( \bm{v}_i - \bm{v}_{\rm c} ) + \bm{v}_{\rm c} , \\[1ex]
\dfrac{\mathrm{d} \bm{v}_i}{\mathrm{d} t} = - \dfrac{1}{\kappa} \dfrac{G m_j (\bm{r}_i - \bm{r}_j)}{|\bm{r}_i - \bm{r}_j|^3} - \dfrac{1}{m_i} \dfrac{\partial V_{\rm ext} (\bm{r})}{\partial \bm{r}_i} ,
\end{cases} \end{align}
where $\bm{v}_{\rm c}$ is the CoM velocity, $V_{\rm ext} (\bm{r})$ is the external potential energy, and $\kappa$ is the SD factor. This factor $\kappa$ slows only the internal motion in the CoM frame, effectively averaging external perturbations over multiple physical orbits and allowing larger timesteps. It is set from the relative strengths of the internal binary interaction and external perturbations \citep{Wang+2020SDAR}:
\begin{align}
\kappa = C \dfrac{m_1 m_2}{(m_1 + m_2) a^3 (1 + e)^3} \sum_{i = 1}^{N_{\rm p}} \dfrac{|\bm{r}_i - \bm{r}_{\rm c}|^3}{m_i} ,
\end{align}
with $C$ a scaling constant, $a$ and $e$ the binary semi-major axis and eccentricity, $N_{\rm p}$ the number of perturbers, and $\bm{r}_{\rm c}$ the CoM position. We also require the slow-down period $T_{\rm SD} = \kappa T$ to satisfy $T_{\rm SD} \leqslant \Delta t$.

Algorithmic regularization is formulated by extending phase space and applying a time transformation. We treat physical time $t$ as an additional coordinate, with conjugate momentum $p_t \equiv -H$. The extended coordinate, momentum, and phase-space vectors are $\bm{R} \equiv (\bm{r}, t)$, $\bm{P} \equiv (\bm{p}, p_t)$, and $\bm{W} \equiv (\bm{R}, \bm{P})$, respectively. We also define the extended kinetic and potential energies as $T(\bm{P}) \equiv T(\bm{p}) + p_t$ and $V(\bm{R}) \equiv V(\bm{r}, t)$. AR introduces a fictitious time $s$ via the time-transformation function $g$:
\begin{align}
g (\bm{W}) = \dfrac{\mathrm{d} t}{\mathrm{d} s} = \dfrac{\ln [T (\bm{P})] - \ln[- V(\bm{R})]}{T (\bm{P}) + V (\bm{R})} .
\end{align}
This choice gives an extended Hamiltonian written in separable logarithmic form,
\begin{align}
\Gamma (\bm{W}) & = g (\bm{W}) [T (\bm{P}) + V (\bm{R})] = \ln T (\bm{P}) - \ln[- V (\bm{R})] ,
\label{eq:Gamma}
\end{align}
which permits explicit symplectic integration, for example with leapfrog (Eq.~\ref{eq:KDK}). The corresponding equations of motion are
\begin{align} \begin{cases}
\dfrac{\mathrm{d} \bm{r}}{\mathrm{d} s} = \dfrac{\partial \Gamma (\bm{R}, \bm{P})}{\partial \bm{p}} = \dfrac{\bm{v}}{T (\bm{p}) + p_t} , \\[2ex]
\dfrac{\mathrm{d} t}{\mathrm{d} s} = \dfrac{\partial \Gamma (\bm{R}, \bm{P})}{\partial p_t} = \dfrac{1}{T (\bm{p}) + p_t} , \\[2ex]
\dfrac{\mathrm{d} \bm{p}}{\mathrm{d} s} = - \dfrac{\partial \Gamma (\bm{R}, \bm{P})}{\partial \bm{r}} = - \dfrac{1}{V (\bm{r}, t)} \dfrac{\mathrm{d} \bm{p}}{\mathrm{d} t} , \\[2ex]
\dfrac{\mathrm{d} p_t}{\mathrm{d} s} = - \dfrac{\partial \Gamma (\bm{R}, \bm{P})}{\partial t} = \dfrac{1}{V (\bm{r}, t)} \dfrac{\partial V (\bm{r}, t)}{\partial t} .
\end{cases} \end{align}

Together, SD and AR efficiently evolve embedded few-body subsystems within larger $N$-body simulations while maintaining accuracy over long times.

\subsubsection{Hamiltonian Splitting through the Particle-Particle Particle-Tree ($\rm P^3 T$) Method} \label{subsubsect:P3T}

We combine the above integrators into a hybrid Hamiltonian-splitting scheme to achieve both high accuracy and efficiency across the full dynamical range of collisional systems. Unlike the standard splitting $H = T + V$, this approach uses the separation of dynamical timescales in collisional $N$-body systems. Most particles evolve smoothly on longer timescales and can be advanced efficiently with lower-order integrators, while a small subset forms compact subsystems with rapid motion requiring higher-order or regularized integration.

More explicitly, we consider an $N$-body system containing $N_\star$ collisional stellar particles and $N - N_\star$ collisionless particles. At each synchronization time, we partition the particle index set $\{ 1, 2, \dots, N \}$ into disjoint subsets. The subsets $\mathcal{G}_k$ ($k = 1, 2, \dots, N_\mathcal{G}$) contain stellar particles assigned to compact subsystems, which we call ``groups'', and the subset $\mathcal{S}$ contains the remaining stellar particles and all collisionless particles, which we call ``singles''. This partition gives
\begin{align}
\mathcal{G}_1 \cup \mathcal{G}_2 \cdots \cup \mathcal{G}_{N_\mathcal{G}} \cup \mathcal{S} = \{ 1, 2, \dots, N \} , \notag \\
\mathcal{G}_k \cap \mathcal{G}_l = \emptyset \quad (\forall~k \neq l) , \quad \mathcal{G}_k \cap \mathcal{S} = \emptyset \quad (\forall~k) .
\label{eq:GS}
\end{align}
The group-identification procedure is described in Sect.~\ref{subsubsect:search}.

Following the P$^3$T-type hybrid method \citep{Oshino2011_P3T, Wang2020PeTar}, \textsc{Arepo-CD} uses this partition to split the Hamiltonian into short- and long-range components. At fixed group membership, this decomposition is algebraically exact:
\begin{align}
H = H_{\rm S} + H_{\rm L} ,
\label{eq:H_split}
\end{align}
with
\begin{align}
H_{\rm S} & = \sum_{i = 1}^N \dfrac{\bm{p}_i^2}{2 m_i} - \sum_{i = 1}^{N_\star - 1} \sum_{j = i + 1}^{N_\star} \dfrac{G m_i m_j}{r_{ij}} W_{ij} ,
\label{eq:H_S} \\
H_{\rm L} & = \sum_{i = 1}^{N_\star - 1} \sum_{j = i + 1}^{N_\star} \dfrac{G m_i m_j}{r_{ij}} \left( W_{ij} - 1 \right) \notag \\
& \quad - \left( \sum_{i = 1}^{N_\star} \sum_{j = N_\star + 1}^{N} + \sum_{i = N_\star + 1}^{N - 1} \sum_{j = i + 1}^{N} \right) G m_i m_j \varphi \left( r_{ij}, \varepsilon_{ij} \right) .
\label{eq:H_L}
\end{align}
Here, $\varphi \left( r_{ij}, \varepsilon_{ij} \right)$ denotes a general softened inverse-distance function with an effective pairwise softening length $\varepsilon_{ij}$, recovering $1 / r_{ij}$ as $\varepsilon_{ij} \to 0$. The pairwise changeover function $W_{ij}$ assigns each interaction to the short- or long-range Hamiltonian. \textsc{Arepo-CD} adopts a sharp changeover based on group membership,\footnote{Conventional P$^3$T implementations generally use a smooth changeover between lower-order tree integration with large shared timesteps and higher-order direct integration with individual timesteps. In \textsc{Arepo-CD}, the long-range tree interactions already follow the individual timestep hierarchy of \textsc{Arepo}, reducing the timestep difference across the transition. We therefore adopt a sharp changeover and assess its accuracy with the tests in Sect.~\ref{sect:test}.} defined as
\begin{align}
W_{ij} = \begin{cases}
1, & i \in \mathcal{G}_k~\text{and}~j \in \mathcal{G}_k \quad (\exists~k) , \\
0, & \text{otherwise} . \end{cases}
\label{eq:W_ij}
\end{align}

The short-range Hamiltonian $H_{\rm S}$ contains the kinetic energy of all particles and the stellar gravitational interactions within each group, while  the long-range Hamiltonian $H_{\rm L}$ contains all remaining stellar interactions and all gravitational interactions involving collisionless particles. Following the operator notation of Eqs.~\ref{eq:KDK} and \ref{eq:FSI}, we denote their corresponding canonical maps over a timestep $\Delta t$ by $E_{\rm S} (\Delta t)$ and $E_{\rm L} (\Delta t)$, respectively. For the disjoint groups $\{ \mathcal{G}_k \}_{k = 1}^{N_\mathcal{G}}$, the short-range map factorizes into independent group integrations and a drift of the single particles:\footnote{This factorization is possible because the factorized Hamiltonians act on disjoint phase-space coordinates and therefore commute.}
\begin{align}
E_{\rm S} (\Delta t) = D_\mathcal{S} (\Delta t) \prod_{k = 1}^{N_\mathcal{G}} E^{(\mathcal{G}_k)} (\Delta t) ,
\label{eq:E_S}
\end{align}
where $D_{\mathcal S}$ drifts the singles and $E^{({\cal G}_k)}$ advances the internal dynamics of group ${\cal G}_k$. The long-range Hamiltonian $H_{\rm L}$ accounts separately for the external gravitational field. Since $H_{\rm L}$ contains only potential terms, its map is a pure kick, $E_{\rm L} = K_{\rm L}$.

We hierarchically couple the three integration methods introduced above through two levels of the short- and long-range Hamiltonian splitting. (1) At the outer level, second-order KDK leapfrog integrates collisionless particles and slowly evolving collisional stellar particles (Sect.~\ref{subsubsect:leapfrog}), and fourth-order FSI evolves the identified stellar groups (Sect.~\ref{subsubsect:4thFSI}). (2) At the inner level, fourth-order FSI evolves most particles within each group, and SDAR integrates the identified tight subgroups (Sect.~\ref{subsubsect:SDAR}). Further details are given below.

Using a symmetric Strang composition, analogous to Eq.~\ref{eq:KDK}, the outer coupling gives the full evolution over one timestep $\Delta t$ as
\begin{align}
E_{\rm KDK} (\Delta t) & \approx K_{\rm L} \left( \dfrac{\Delta t}{2} \right) E_{\rm S}(\Delta t) K_{\rm L} \left( \dfrac{\Delta t}{2} \right) \notag \\
& = K_{\rm L} \left( \dfrac{\Delta t}{2} \right) D_\mathcal{S} (\Delta t) \prod_{k = 1}^{N_\mathcal{G}} E_{\rm FSI+SDAR}^{(\mathcal{G}_k)} (\Delta t) K_{\rm L} \left( \dfrac{\Delta t}{2} \right) .
\label{eq:KDK_P3T}
\end{align}
Interactions between stellar groups and single particles outside them are handled by the long-range kicks $K_{\rm L}$, evaluated using the gravitational force octree (Sect.~\ref{subsect:overview}; \citealt{Springel2005GADGET-2, Springel2010Arepo, Weinberger+2020Arepo, Springel2021GADGET-4}). Single particles outside the groups retain the standard KDK evolution without group integration.

To balance accuracy and computational efficiency within each stellar group, we identify tight subgroups, such as close binaries and hierarchical triples, using a partition similar to Eq.~\ref{eq:GS}. Fourth-order FSI integrates the general few-body dynamics, while SDAR handles the stiff subgroups more efficiently. For compactness, we define the combined SDAR evolution of all tight subgroups within group $\mathcal G_k$ as
\begin{align}
R_{\rm SDAR}^{(\mathcal{G}_k)} (\Delta t) \equiv \prod_{l = 1}^{N_{\mathcal F}^{(\mathcal G_k)}} E_{\rm SDAR}^{(\mathcal F_l)} (\Delta t) ,
\end{align}
where $\{ \mathcal{F}_l \}_{l = 1}^{N_\mathcal{F}^{(\mathcal{G}_k)}}$ denotes the set of tight subgroups identified within $\mathcal{G}_k$, and the map $E_{\rm SDAR}^{(\mathcal F_l)} (\Delta t)$ advances subgroup $\mathcal F_l$ with SDAR over one timestep $\Delta t$. The full evolution map of group $\mathcal G_k$ over $\Delta t$ is then
\begin{align}
E_{\rm FSI+SDAR}^{(\mathcal{G}_k)} (\Delta t) = &~ K_{\rm L}^{(\mathcal{G}_k)} \left( \dfrac{\Delta t}{6} \right) D_{\rm S}^{(\mathcal{G}_k)} \left( \dfrac{\Delta t}{2} \right) R_{\rm SDAR}^{(\mathcal{G}_k)} \left( \dfrac{\Delta t}{2} \right) \tilde{\mathcal{K}}_{\rm L}^{(\mathcal{G}_k)} \left( \dfrac{2}{3} \Delta t \right) \notag \\
&~ D_{\rm S}^{(\mathcal{G}_k)} \left( \dfrac{\Delta t}{2} \right) R_{\rm SDAR}^{(\mathcal{G}_k)} \left( \dfrac{\Delta t}{2} \right) K_{\rm L}^{(\mathcal{G}_k)} \left( \dfrac{\Delta t}{6} \right),
\label{eq:FSI+SDAR}
\end{align}
where $K_{\rm L}^{(\mathcal G_k)}$ and $D_{\rm S}^{(\mathcal G_k)}$ denote the kick and drift operators, respectively, acting on the remaining single stellar particles within group $\mathcal G_k$. The operator $\tilde{\mathcal{K}}_{\rm L}^{(\mathcal G_k)}$ is the corresponding gradient-corrected kick. For sufficiently small groups, evolving the entire group with SDAR is more efficient. We therefore use SDAR alone when $|\mathcal{G}_k| \leqslant N_{\rm SDAR, max}$. Following \citet{Wang+2020SDAR}, we typically set $N_{\rm SDAR, max} \leqslant 4$.

Substituting the compact stellar group integration of Eq.~\ref{eq:FSI+SDAR} into the global time evolution of Eq.~\ref{eq:KDK_P3T} yields the three-level evolution operator that forms the core integration algorithm of \textsc{Arepo-CD}. The combined scheme is formally second-order accurate because the outermost coupling uses a second-order KDK leapfrog integrator. Its numerical accuracy is also affected by the approximate gravitational forces from the octree. Nevertheless, the few-body dynamics are treated with higher-order integrators, while the three-level structure covers a wide dynamical range. The benchmark tests in Sect.~\ref{sect:test} show that \textsc{Arepo-CD} agrees well with $N$-body calculations and reproduces the key dynamical properties of star clusters.

\subsubsection{Parallel Group Integration} \label{subsubsect:parallel}

Next, we provide further details on the integration of the compact stellar groups $\{ \mathcal{G}_k \}_{k = 1}^{N_\mathcal{G}}$ defined in Sect.~\ref{subsubsect:P3T}. For each group, we compute its CoM properties and transform its members’ positions and velocities into the CoM frame. This separates the CoM motion from the internal motion. The CoM motion, together with that of single stellar and collisionless particles, is advanced with the standard KDK leapfrog scheme (Eq.~\ref{eq:KDK}), while the internal relative dynamics are evolved with the fourth-order FSI+SDAR solver (Eq.~\ref{eq:FSI+SDAR}). The global timestep is determined from active particles outside the groups together with the group CoM contributions. All members of a group are assigned the same global timestep as their CoM to maintain synchronization with the global KDK integration. The above approach is also applied to subgroups integrated with SDAR within each compact stellar group.

The decomposition into disjoint group integrations in Eq.~\ref{eq:KDK_P3T} also allows independent group integration in parallel across MPI ranks, as external gravitational interactions are handled separately through the long-range kick $K_{\rm L}$ rather than during internal group integration. \textsc{Arepo-CD} distributes the groups with a greedy load-balancing scheme based on an approximate integration-cost estimate following \citet{Wang+2020SDAR}. Because only group-member data are exchanged during this stage, the communication overhead between MPI ranks remains minimal. After integration, the updated particle properties are returned to the MPI ranks on which the corresponding particles reside.

As an accuracy check, the total internal energy $E$ of each group is recomputed at the end of the integration. If the relative energy error exceeds a predefined threshold, the ICs of the group are written to a debug file for further inspection. An optional configuration also records the cumulative integration error over all groups for performance diagnostics and parameter tuning.

\subsection{Timestep Criteria} \label{subsect:timestep}

We introduce timestep criteria for collisional stellar particles to resolve short dynamical timescales accurately \citep{Dehnen&Read2011NBody} and to support the group identification in Sect.~\ref{subsect:group}. The standard \textsc{Arepo} gas timestep criteria already account for the relevant hydrodynamic and MHD constraints and are therefore left unchanged \citep{Springel2010Arepo, Weinberger+2020Arepo}.

For stellar particles, however, the default gravitational timestep is only controlled by the acceleration criterion commonly used in cosmological simulations \citep{Power2003_accel},
\begin{align}
\Delta t \leqslant \eta_{\rm accel} \sqrt{\dfrac{\varepsilon}{|\bm{a}|}} .
\label{eq:timestep_accel}
\end{align}
Here $\varepsilon$ is a characteristic length scale, which we set to the gravitational softening length. The particle acceleration is $\bm{a}$ and $\eta_{\rm accel}$ is a dimensionless safety parameter. Although this criterion is effective in many collisionless applications, it is not sufficiently robust for the collisional stellar dynamics followed in \textsc{Arepo-CD}. Close binaries, higher-order multiples, and strong encounters can change stellar velocities on very short timescales \citep{Valtonen&Karttunen2006ThreeBody}, which are not always captured by an acceleration criterion alone. Moreover, geometric cancellation can also reduce the magnitude of the instantaneous acceleration $\bm{a}$, which permits excessively large timesteps in dynamically important regions (see Fig.~1 of \citealt{Grudic&Hopkins2020tidal}). Robust and efficient timestep control in collisional star-gas simulations therefore requires additional conservative criteria for stellar particles. We introduce these criteria below.

\subsubsection{Tidal Criterion} \label{subsect:tidal}

We first add the tidal timestep criterion of \citet{Grudic&Hopkins2020tidal} to avoid failures caused by geometric cancellation of the acceleration. This criterion estimates the local dynamical timescale from the tidal tensor $\bm{T} = \nabla \bm{a}$, i.e., the spatial gradient of the gravitational acceleration. The invariant Frobenius norm of $\bm{T}$ gives
\begin{align}
\Delta t & \leqslant \eta_{\rm tidal} t_{\rm tidal} = \eta_{\rm tidal} \left( \dfrac{||\bm{T}||^2}{6} \right)^{- 1 / 4} \notag \\
& = \eta_{\rm tidal} \left( \dfrac{\sum_{i = 1}^3 \sum_{j = 1}^3 T_{ij}^2}{6} \right)^{- 1 / 4} ,
\end{align}
which recovers the usual Keplerian dynamical timescale in the appropriate limit. Here $\eta_{\rm tidal}$ is a dimensionless safety factor. When a softened force law is used, the tidal tensor must be computed consistently from the softened gravitational potential \citep{Hopkins2015GIZMO, Grudic+2021STARFORGE}.

\subsubsection{Free-fall and Fly-by Criteria} \label{subsubsect:ff&fb}

To capture the short timescales associated with sudden velocity changes during close few-body interactions, we supplement the timestep control with two pairwise criteria based on physically motivated two-body timescales \citep{Rantala+2021FROST, Rantala+2023BIFROST}. For each stellar particle $i$, we consider neighboring stellar particles $j$ within the search radius $r_{\rm search}$, an adjustable input parameter.\footnote{The search radius is also adjusted adaptively according to the particle configuration to avoid prohibitively expensive $\mathcal{O}(N^2)$ pairwise calculations.} The two criteria use the pairwise free-fall timescale $\Delta t_{{\rm ff}, ij}$ and fly-by timescale $\Delta t_{{\rm fb}, ij}$, respectively. We take the minimum over neighboring stars within $r_{\rm search}$ and impose
\begin{align}
\Delta t \leqslant \eta_{\rm ff} \min_{\substack{j \neq i, \\ r_{ij} < r_{\rm search}}} \Delta t_{{\rm ff}, ij} = \eta_{\rm ff} \min_{\substack{j \neq i, \\ r_{ij} < r_{\rm search}}} \sqrt{\dfrac{r_{ij}^3}{G (m_i + m_j)}} ,
\label{eq:ff}
\end{align}
and
\begin{align}
\Delta t \leqslant \eta_{\rm fb} \min_{\substack{j \neq i \\ r_{ij} < r_{\rm search}}} \Delta t_{{\rm fb}, ij} = \eta_{\rm fb} \min_{\substack{j \neq i \\ r_{ij} < r_{\rm search}}} \dfrac{r_{ij}}{v_{ij}} .
\label{eq:fb}
\end{align}
Here $r_{ij} = |\bm{r}_{ij}| = |\bm{r}_i - \bm{r}_j|$ is the pairwise separation, $v_{ij} = |\bm{v}_{ij}| = |\bm{v}_i - \bm{v}_j|$ is the relative speed, and $m_{ij} = m_i + m_j$ is the total mass. The dimensionless safety factors $\eta_{\rm ff}$ and $\eta_{\rm fb}$ are specified as input parameters. In practice, we recommend $\eta_{\rm ff} = \eta_{\rm fb} \approx 2~\eta_{\rm tidal} \sim 0.02$, so that the two-body criteria then primarily control the timestep, and the tidal criterion provides an additional safeguard.

We note that, to resolve close stellar neighbors, all tree nodes that overlap the search region of radius $r_{\rm search}$ are opened. We also reuse this neighbor information for timestep calculations to identify groups during the gravity octree walk (Sect.~\ref{subsubsect:pair}).

\subsection{Group and Subgroup Identification} \label{subsect:group}

We now describe how \textsc{Arepo-CD} identifies compact stellar ``groups'' (defined by Eq.~\ref{eq:GS}) for integration with the few-body solvers in Sect.~\ref{subsubsect:P3T}. At the beginning of each timestep, the algorithm (1) identifies close stellar ``pairs'' (defined below) during the gravity calculation using the Barnes--Hut octree and then (2) merges these pairs hierarchically and adaptively into disjoint groups. The same procedure is used to identify subgroups within each group, except that the Barnes--Hut octree is replaced by direct force calculations. The following subsections therefore describe both stages using group identification as an example, with subgroup-specific details given where necessary.

\subsubsection{Pair Identification} \label{subsubsect:pair}

During the gravity octree walk, the pairwise free-fall and fly-by timescales, $\Delta t_{{\rm ff}, ij}$ and $\Delta t_{{\rm fb}, ij}$ (defined in Eqs.~\ref{eq:ff} and \ref{eq:fb}), 
are reused for group identification. Two stellar particles $(i, j)$ are identified as a collisional ``pair'' if either characteristic interaction timescale is sufficiently short. Specifically, we require
\begin{align}
\Delta t_{{\rm ff}, ij} = \sqrt{\dfrac{r_{ij}^3}{G (m_i + m_j)}} < \Delta t_{\rm group} ,
\label{eq:pair_ff}
\end{align}
or
\begin{align}
\Delta t_{{\rm fb}, ij} = \dfrac{r_{ij}}{v_{ij}} < \Delta t_{\rm group} .
\label{eq:pair_fb}
\end{align}
Here $r_{ij}$, $v_{ij}$, and $m_i + m_j$ have the same definitions as in Eqs.~\ref{eq:ff} and \ref{eq:fb} and we require $j \neq i$ and $r_{ij} < r_{\rm search}$. The input parameter $\Delta t_{\rm group}$ identifies stellar particles with short dynamical timescales that require specialized few-body integration, with a typical value of $\sim 10~\rm yr$.

The pair identification is performed during the timestep calculation in the gravity force-tree walk, avoiding an additional stellar neighbor search. Moreover, as discussed in Sect.~\ref{subsubsect:parallel}, stellar particles assigned to groups no longer directly impose their short internal timescales on the global integration. The few-body solvers evolve these particles on their short internal timescales, allowing a larger global timestep and improving the overall integration efficiency.

\subsubsection{Adaptive Group Construction} \label{subsubsect:search}

\begin{figure}[htbp!]
\centering
\includegraphics[width=\linewidth]{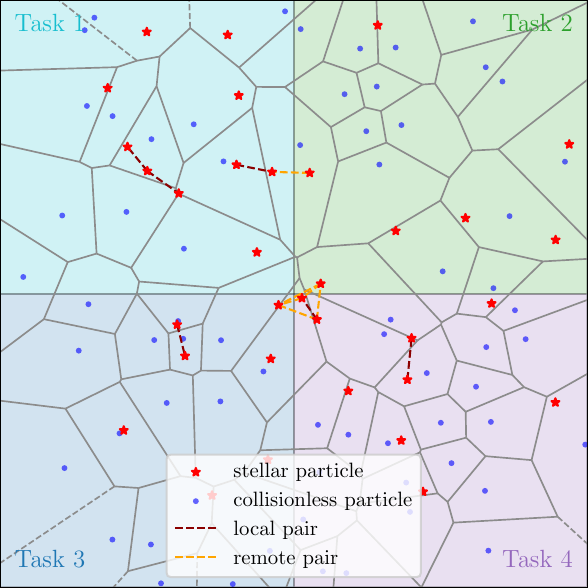}
\caption{Schematic illustration of the group search across four MPI tasks. Red stars denote collisional stellar particles and blue dots denote collisionless particles. Dashed lines connect stellar pairs identified during the gravity calculation. A ``local pair'' (dark red) contains two stars on the same MPI task, whereas a ``remote pair'' (orange) connects stars on different tasks. For each identified pair, the two stars belong to the same collisional group. If either star also forms a pair with another star, that star is added to the same group. This procedure is repeated until all such stars are assigned to the group.}
\label{fig:group}
\end{figure}

As Fig.~\ref{fig:group} illustrates, the identified collisional pairs are used to construct groups in a friends-of-friends (FOF) way \citep{Hopcroft&Tarjan1973graph, Rantala+2023BIFROST, Bernard+2025PhantomN}. The two stars in each pair are assigned to the same collisional group. If either star forms another pair with a third star, the third star is added to the same group. This procedure is implemented using a disjoint-set data structure and the union--find algorithm, and is repeated until no additional stars can be added. The group search first identifies groups for the fourth-order FSI, and then identifies smaller subgroups within each FSI group that require SDAR integration. The first stage is parallelized across MPI tasks, whereas the second is performed locally because all members of an FSI group are already stored on the same CPU. For the parallel search, we distinguish between ``local pairs'', whose two members reside on the same MPI task, and ``remote pairs'', whose members reside on different tasks. Each task first constructs groups from its local pairs. The tasks then exchange remote pairs and use all pairs to merge groups across task boundaries, yielding the final disjoint groups and ungrouped single stars.

This approach provides an alternative to the default FOF algorithm in \textsc{Arepo} with lower overhead \citep{Springel2010Arepo, Weinberger+2020Arepo}. The standard \textsc{Arepo} FOF method requires an additional tree containing only stars and a separate tree traversal, introducing extra memory and computational costs. By contrast, our approach identifies stellar pairs directly during the gravity calculation. 
The pairs must be stored explicitly, but the additional memory cost is modest because the collisional groups considered here are much smaller than the halo-scale FOF groups commonly used in cosmological simulations \citep[e.g.,][]{Davis+1985FOF}. For example, even if all stellar particles belong to isolated binary stars, the memory cost remains $\mathcal{O} (N)$. In practice, the group-search cost is negligible compared with the subsequent group integration. For an extremely compact cluster, the memory required to store pairs can approach $\mathcal{O} (N^2)$, although this can be alleviated by the adaptive group-search strategy described below.

More importantly, such rare but prohibitively expensive groups also affect the group integration, since the largest group usually dominates the integration runtime \citep{Wang+2015NBody6++GPU, Wang2020PeTar}. The strategy is to reduce $\Delta t_{\rm group}$ to form smaller groups, together with an explicit upper limit on the size of each collisional group. Exceptionally large groups can occasionally exceed $\sim 10^3$ members for the fourth-order FSI or $\sim 10$ members for SDAR, making the integrators inefficient. During the union--find construction, we therefore track the size of each group and process pairs in increasing order of their minimum dynamical timescale, $\min \left\{ \Delta t_{\mathrm{ff}, ij}, \Delta t_{\mathrm{fb}, ij} \right\}$, with $j \neq i$ and $r_{ij} < r_{\rm search}$. A merger is allowed only if the resulting group or subgroup remains below the user-defined limits, $N_{\rm FSI,max}$ (typically $\sim 100$) for FSI groups or $N_{\rm SDAR,max}$ for the SDAR subgroups (introduced in Sect.~\ref{subsubsect:P3T}). If a merger would exceed the group-size limit, it is skipped and the remaining candidate pairs are processed. Interactions excluded from group integration remain subject to the timestep criteria in Sect.~\ref{subsect:timestep}, which can then reduce the global timestep accordingly. This trades smaller timesteps for a predictable computational cost of the embedded (sub)group integration using the (FSI+)SDAR solver.

\subsection{Coupling to \textsc{Rigel}} \label{subsect:RIGEL}

To model individual star formation and feedback during star cluster formation and early evolution in a realistic gaseous environment, we couple \textsc{Arepo-CD} to \textsc{Rigel} \citep{Deng+2024RIGEL, Deng+2026aRIGEL}, a star-by-star galaxy formation framework built on \textsc{Arepo}. The gas is evolved with the moving-mesh hydrodynamics and ideal-MHD solvers of \textsc{Arepo}, while radiation is coupled through the moment-based \textsc{Arepo-RT} solver \citep{2019MNRAS.485..117K, 2024MNRAS.533..268Z:Zier}. The model includes detailed interstellar medium (ISM) cooling, \textsc{Hii}-region feedback, and supernova (SN) feedback \citep{2023ApJS..264...10K, 2024MNRAS.527..478D, 2015ApJ...802...99K}. Stellar lifetimes, photon production rates, mass-loss rates, and wind velocities are determined from the initial mass and metallicity of each star using the stellar-model library described in \citet{Deng+2024RIGEL}. An updated individual-star formation model \citep{Deng+2026aRIGEL} samples stars over the full stellar initial mass function (IMF) from $0.08$ to $150~M_\odot$ and reproduces the observed maximum stellar mass--cluster mass relation. Further details of the \textsc{Rigel} framework are given in \citet{Deng+2024RIGEL, Deng+2025RIGEL, Deng+2026aRIGEL, Deng+2026bRIGEL}.

\textsc{Arepo-CD} and \textsc{Rigel} are coupled flexibly, allowing both the \textsc{Rigel} model and collisional dynamics in \textsc{Arepo} to be switched on or off independently for the tests presented in Sects.~\ref{sect:test} and \ref{sect:hydro}. Without \textsc{Arepo-CD}, stellar gravity is computed with the Barnes--Hut octree \citep{Barnes&Hut1986octree} and gravitational softening \citep{Springel2010Arepo, Weinberger+2020Arepo}. Enabling collisional dynamics also involves a modification to the initial conditions of newly formed stars. The original \textsc{Rigel} model randomly offsets the initial position of each newly formed star from that of its parent reservoir particle, which represents the protostellar phase through a subgrid model (see \citealt{Deng+2026aRIGEL} for details). We additionally introduce a random velocity perturbation to reduce the formation of unphysically compact primordial binaries. For each Cartesian coordinate, the velocity offset is drawn independently from a zero-mean Gaussian distribution with the Box--Muller method. The newly introduced velocity dispersion $\sigma_v$ is set to $\sim 0.1~{\rm km/s}$ in our fiducial run to minimize its effect on the final stellar velocities.

Although the \textsc{Rigel} framework models individual stars, resolving full collisional dynamics for every star can be prohibitively expensive in galaxy-scale simulations \citep{Lahen+2025aSPHGAL+KETJU, Lahen+2025bSPHGAL+KETJU}. We therefore provide an option that enables collisional dynamics only in regions around massive stars (e.g. $> 8~M_\odot$, with the threshold treated as a free parameter). This option reduces the computational cost and focuses the collisional calculation on massive stars. We disable this option in the production runs presented here because these runs cover only GMC and star cluster scales.

\section{Benchmark Tests} \label{sect:test}

We perform a suite of controlled $N$-body benchmark tests to assess the numerical methods described in Sect.~\ref{sect:method}. We compare \textsc{Arepo-CD} with the collisional $N$-body code \textsc{PeTar}\footnote{GitHub repository: \url{https://github.com/lwang-astro/PeTar}}, a code developed for collisional stellar systems \citep{Iwasawa2016_FDPS, Namekata2018_FDPS, Wang2020PeTar}. We use two complementary tests. The first compares the collisional evolution with \textsc{PeTar} and theoretical expectations, while the second isolates the effects of collisional dynamics through comparison with simulations that use softened gravity. Throughout this work, simulations performed with \textsc{PeTar} are referred to as ``\textsc{PeTar}'', those with \textsc{Arepo} including collisional dynamics as ``\textsc{Arepo-CD}'', and those with softened gravity as ``\textsc{Arepo}''.

The initial conditions (ICs) are isolated, spherically symmetric star cluster models generated with a new version of the \textsc{McLuster} code\footnote{GitHub repository: \url{https://github.com/lwang-astro/mcluster}.} \citep{Kupper+2011McLuster, Wang+2019McLuster}. We adopt the \citet{Plummer1911} sphere for the initial density profile and initialize the systems in virial equilibrium. For simplicity, we neglect external tidal fields, stellar evolution, and primordial binaries. Sects.~\ref{subsect:radii} and \ref{subsect:rho&vel} give further details of the two simulation setups.

\subsection{Core and Lagrangian Radii} \label{subsect:radii}

\begin{figure}[htbp!] \centering
\includegraphics[width=\linewidth]{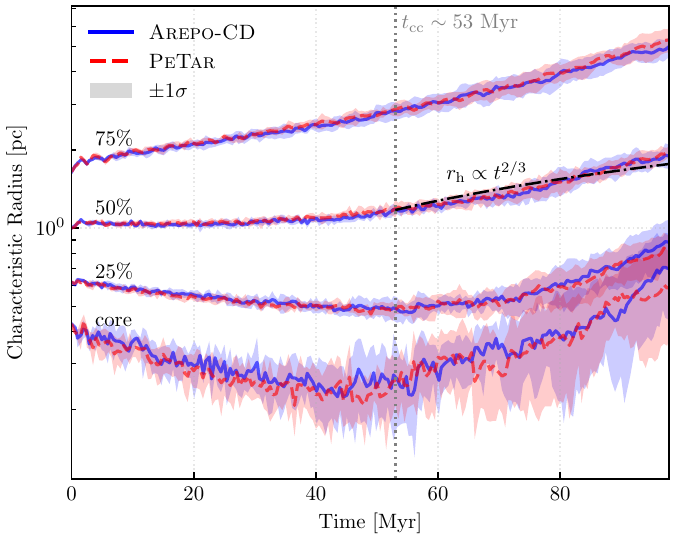}
\caption{Evolution of the core radius and Lagrangian radii for the $N = 10^3$ equal-mass star cluster, with the ICs described in the main text. Blue solid curves show \textsc{Arepo-CD}, while red dashed curves show \textsc{PeTar}. From bottom to top, the curves correspond to the core radius and the 25\%, 50\%, and 75\% Lagrangian radii. Shaded regions indicate the $\pm 1 \sigma$ scatter from eight random realizations. The gray dotted vertical line marks the expected core-collapse time, $t_{\rm cc} \sim 53~{\rm Myr}$. The black dash-dotted line shows the expected self-similar expansion after core collapse, $r_{\rm h} \propto t^{2/3}$. Overall, \textsc{Arepo-CD} agrees well with \textsc{PeTar} and reproduces the theoretically expected core-collapse and subsequent evolution.}
\label{fig:log_radii}
\end{figure}

\begin{figure*}[htbp!] \centering
\begin{subfigure}{0.49\textwidth} \centering
\includegraphics[width=\linewidth]{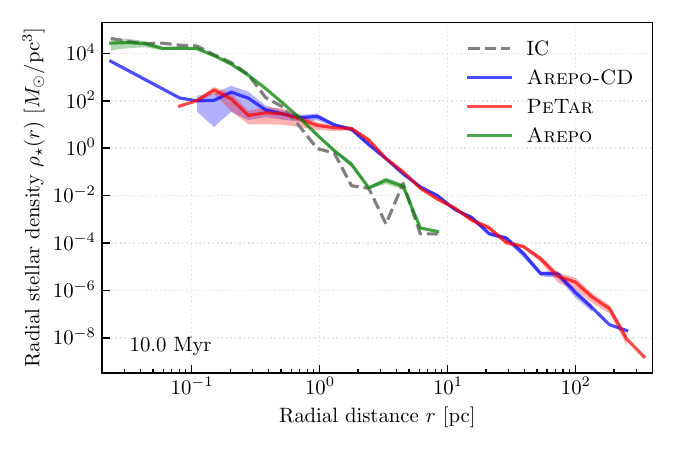}
\end{subfigure} \hfill \begin{subfigure}{0.49\textwidth} \centering
\includegraphics[width=\linewidth]{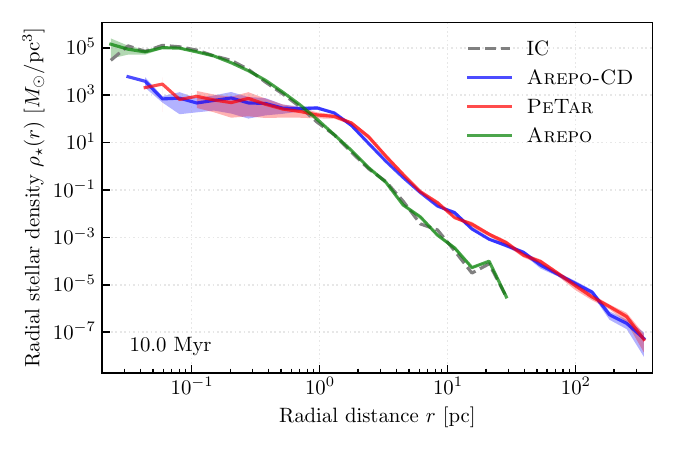} \end{subfigure}
\\
\begin{subfigure}{0.49\textwidth} \centering
\includegraphics[width=\linewidth]{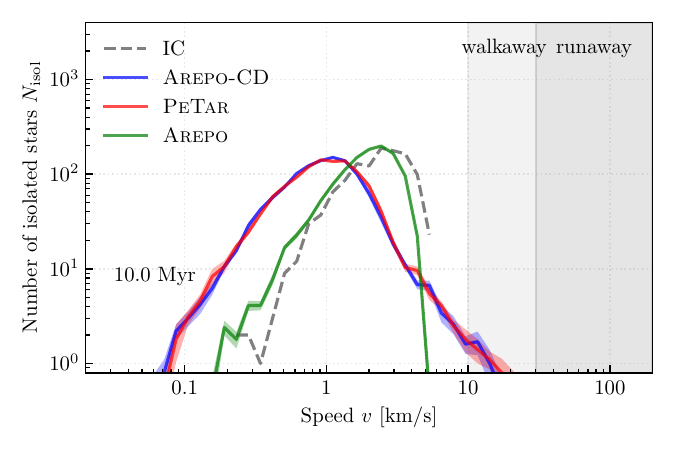}
\end{subfigure} \hfill \begin{subfigure}{0.49\textwidth} \centering
\includegraphics[width=\linewidth]{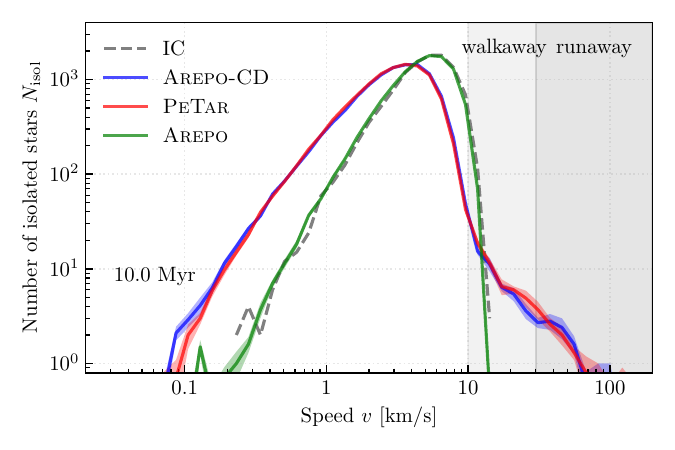} \end{subfigure}
\caption{Radial density profiles (top) and velocity distributions (bottom) of isolated star clusters with realistic IMFs at $t = 10~\rm Myr$, containing $N_\star \sim 10^3$ (left) and $\sim 10^4$ (right) stars. The velocity distributions include only isolated stars with no neighbor within $0.01~\rm pc$. Dashed gray curves show the ICs, while the blue, red, and green curves correspond to the \textsc{Arepo-CD}, \textsc{PeTar}, and \textsc{Arepo} runs, respectively. Each curve shows the mean over 10 random realizations, with the colored shading indicating the corresponding $\pm 1 \sigma$ scatter. Gray shaded regions mark the walkaway ($10~{\rm km/s} < v \leqslant 30~{\rm km/s}$) and runaway ($v > 30~{\rm km/s}$) regimes. \textsc{Arepo-CD} closely reproduces the \textsc{PeTar} results in both the radial density profiles and velocity distributions, including the high-velocity escaper tail. The \textsc{Arepo} runs with softened stellar gravity remain more centrally concentrated and produce fewer fast escapers.}
\label{fig:1k_10k}
\end{figure*}

We first compare \textsc{Arepo-CD} with \textsc{PeTar} using an idealized equal-mass star cluster whose theoretical evolution is well established. We use a cluster with $N = 10^3$ stars, each with a mass of $m_\star = 8~M_\odot$.\footnote{This choice of stellar mass is a convenient rescaling that allows the cluster to be evolved over $\sim 100~\rm Myr$. It does not affect the underlying dimensionless dynamics.} The initial half-mass radius is $r_{\rm h, 0} = 1.0~{\rm pc}$. To quantify stochastic variation, we generate eight random realizations of the ICs for each code.

For this model, the initial half-mass relaxation time is estimated as
\begin{align}
t_{\rm hr} = \dfrac{0.138}{\ln \Lambda} \sqrt{\dfrac{N r_{\rm h}^3}{G m_\star}} \simeq 3.8~{\rm Myr} ,
\label{eq:hr}
\end{align}
where we adopt the conventional Coulomb logarithm $\ln \Lambda = \ln (0.4 N)$ \citep{Spitzer1987}. For an isolated equal-mass Plummer model, core collapse occurs after $\sim14$--$17 t_{\rm hr}$ \citep{Cohn1980N-body, Giersz&Heggie1994N-body, Makino1996N-body}, corresponding to $t_{\rm cc} \sim 53$--$65~{\rm Myr}$ here. Hard-binary formation then halts core collapse and drives post-collapse expansion. The half-mass radius approaches the scaling $r_{\rm h} \propto t^{2/3}$ \citep{Giersz&Heggie1994N-body, Henon+2011N-body}.

We now compare these expectations with the numerical evolution. Fig.~\ref{fig:log_radii} compares the core radius and the Lagrangian radii enclosing 25\%, 50\%, and 75\% of the bound stellar mass. The core radius and 25\% Lagrangian radius probe the central collisional dynamics, while the 50\% and 75\% radii trace the global mass distribution and secular expansion. Before core collapse, the core contracts while the half-mass radius remains nearly constant as expected from two-body relaxation \citep{Giersz&Heggie1994N-body}. \textsc{Arepo-CD} is statistically consistent with \textsc{PeTar} throughout the evolution. Both show the expected core collapse at $t_{\rm cc} \sim 53~{\rm Myr}$ and subsequent self-similar expansion, $r_{\rm h} \propto t^{2/3}$. The agreement across all radii shows that \textsc{Arepo-CD} reproduces both the inner collisional evolution and the global relaxation behavior, with residual differences smaller than the variation between realizations.

\subsection{Radial Density Profile and Velocity Distribution} \label{subsect:rho&vel}

For the second test, we consider isolated young star clusters with a realistic stellar IMF to isolate the effects of collisional dynamics. Individual stellar masses are drawn from the \cite{Kroupa2001IMF} IMF, with the maximum stellar mass set from the cluster mass following \cite{Yan+2023YSC}. See also \cite{Lahen+2025aSPHGAL+KETJU} and \cite{Deng+2026aRIGEL}. The initial three-dimensional half-mass radius is assigned from the observed cluster mass-radius relation \citep{Marks&Kroupa2012YSC}:
\begin{align}
\dfrac{r_{\rm h, 0}}{\rm pc} = \dfrac{f_{\rm h} R_4}{1.3} \left( \dfrac{M_\star}{10^4~M_\odot} \right)^\beta .
\label{eq:mass-radius}
\end{align}
Here we adopt $\beta = 0.18 \pm 0.028$ and $R_4 = 2.365 \pm 0.106$ from \cite{Brown&Gnedin2021YSC}, together with $f_{\rm h} \sim 1 / 6$ to represent the compact birth radii inferred for embedded clusters observationally \citep[e.g.,][]{Adamo+2024}. Recent dynamical models also adopt such compact radii \citep[e.g.,][]{Rantala+2024FROST, Rantala+2025FROST, Rantala+2026FROST}. We study two models containing $\sim 10^3$ and $\sim 10^4$ stars, with total stellar masses of $\sim 600$ and $\sim 6000~M_\odot$, half-mass radii of $0.18$ and $0.28~\rm pc$, and maximum stellar masses of $\sim 15$ and $\sim 55~M_\odot$, respectively. These compact clusters have short crossing and relaxation times and are therefore expected to undergo substantial structural evolution within $\sim 10~\rm Myr$. For each model, we generate 10 random realizations.

Fig.~\ref{fig:1k_10k} compares the stellar radial density profiles and velocity distributions at $t = 10~\rm Myr$. The top two panels show the radial density profiles of the two cluster models. The \textsc{Arepo} runs with softened stellar gravity remain close to the ICs. They retain a centrally concentrated stellar distribution and develop only a weak extended tail. In contrast, both \textsc{Arepo-CD} and \textsc{PeTar} deviate substantially from the ICs and the \textsc{Arepo} results with softened gravity, showing reduced central densities and extended outer density tails. \textsc{Arepo-CD} agrees well with \textsc{PeTar} over most of the resolved radial range. Larger stochastic differences occur mainly in the innermost bins, where only a small number of stars are present and strong few-body encounters dominate. This evolution is consistent with collisional dynamics. Two-body relaxation and mass segregation drive energy exchange in the core, while hard-binary and few-body interactions heat the cluster and move stars to larger radii \citep{Heggie&Hut2003N-body}.

Next, we compare the velocity distributions in the bottom two panels of Fig.~\ref{fig:1k_10k}. We consider only stars with no neighbor within $0.01~\rm pc$. This selection excludes large instantaneous velocities caused by internal binary motion and focuses on escapers. We classify stars with $10~{\rm km/s} < v \leqslant 30~{\rm km/s}$ as walkaways and those with $v > 30~{\rm km/s}$ as runaways, with these ranges marked by gray shaded regions \citep{Eldridge2011runaway&BSS&walkaway, Schoettler+2019runaway&DES&walkaway}. Their origin is discussed in Sect.~\ref{subsect:vel}. The softened \textsc{Arepo} runs produce fewer high-velocity escapers, particularly in the more massive cluster. In contrast, \textsc{Arepo-CD} and \textsc{PeTar} show nearly identical speed distributions, including the high-velocity tail generated by strong collisional interactions.

\section{Self-consistent simulations of star cluster formation and dynamical evolution} \label{sect:hydro}

\begin{figure*}[htbp!] \centering
\begin{subfigure}{0.49\textwidth} \centering
\includegraphics[width=\linewidth]{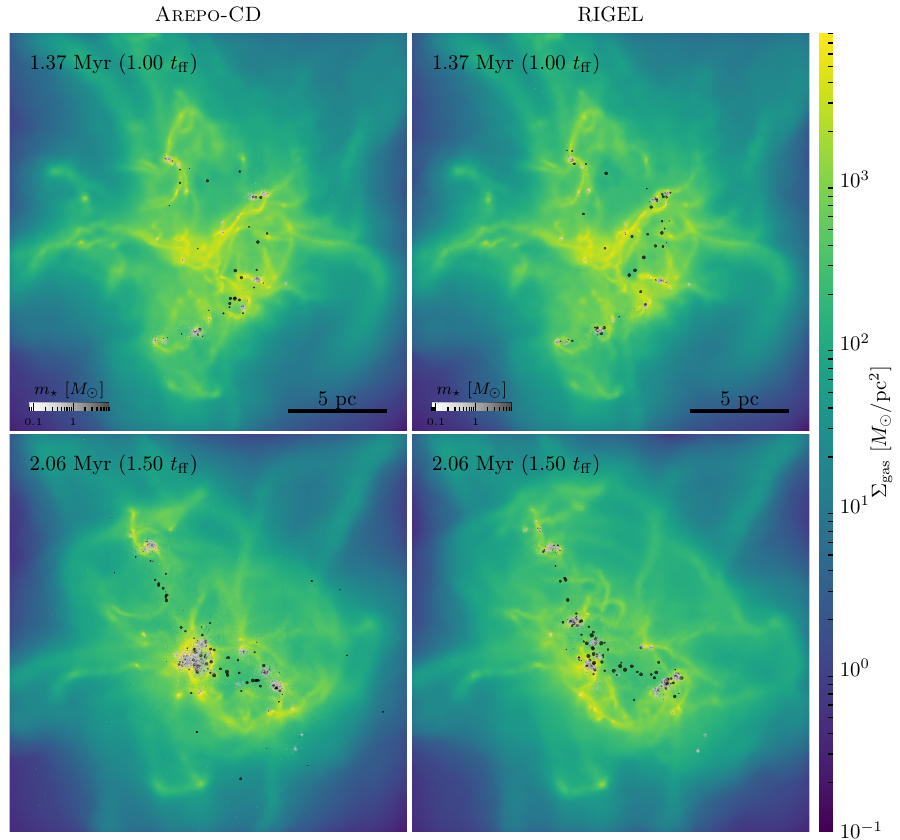}
\end{subfigure} \hfill \begin{subfigure}{0.49\textwidth} \centering
\includegraphics[width=\linewidth]{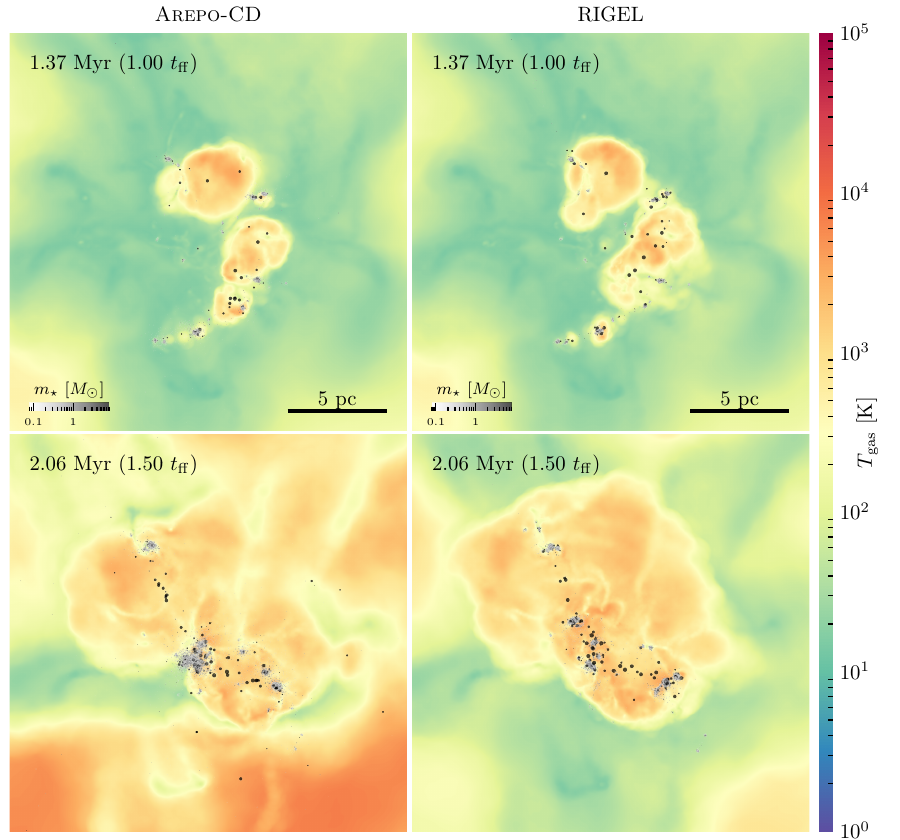} \end{subfigure}
\caption{Comparison of the collisional \textsc{Arepo-CD} and collisionless \textsc{Rigel} GMC simulations. The top row shows $t \simeq 1.37~{\rm Myr}$ ($\simeq 1.0~t_{\rm ff}$) and the bottom row shows $t \simeq 2.06~{\rm Myr}$ ($\simeq 1.5~t_{\rm ff}$). The left pair of columns shows the projected gas surface density, $\Sigma_{\rm gas}$, for \textsc{Arepo-CD} and \textsc{Rigel}, while the right pair shows the corresponding projected mass-weighted gas temperature, $T_{\rm gas}$. Stellar particles are overlaid and colored by stellar mass. All panels use the same field of view, with a 5 pc scale bar. At $t \simeq 1.0~t_{\rm ff}$, the two runs remain nearly indistinguishable in both their large-scale gas structure and stellar distribution. By $t \simeq 1.5~t_{\rm ff}$, collisional dynamics has redistributed the massive stars in \textsc{Arepo-CD}, producing clear differences in the locations and shapes of the heated regions.}
\label{fig:Sigma_T}
\end{figure*}

Next, we turn to the full hydrodynamical GMC simulations that follow the formation and early evolution of star clusters. We analyze a fiducial model in which an isolated, isothermal, massive, turbulent GMC collapses to form stars, allowing us to characterize both the new $N$-body scheme and the resulting stellar system. Throughout this section, ``\textsc{Rigel}'' denotes simulations with the \textsc{Rigel} model described in Sect.~\ref{subsect:RIGEL}, while ``\textsc{Arepo-CD}'' denotes simulations with \textsc{Rigel} and collisional dynamics.

\subsection{Initial Conditions and Simulation Parameters} \label{subsect:IC}

The simulations start from an isolated, isothermal, turbulent GMC generated with \textsc{MakeCloud}\footnote{GitHub repository: \url{https://github.com/mikegrudic/makecloud}} \citep{Grudic&Guszejnov2021MakeCloud}, following a setup similar to \cite{Bernard+2025PhantomN}. The cloud has mass $M_{\rm GMC} = 10^5~M_\odot$ and radius $R_{\rm GMC} = 8.8~\rm pc$, embedded in an ambient medium with density $10^{-3}$ times that of the cloud. We place the cloud at the center of a periodic box with side length $\sim 200~\rm pc$. The gas is initialized at $T = 20~\rm K$ with mean molecular weight $\mu = 2.35$. We add a Gaussian random velocity field with power spectrum $P_k \propto k^{-2}$, scaled to a turbulent virial ratio $\alpha_{\rm turb} = 2$. We use a gas mass resolution of $m_{\rm gas} \simeq 1~M_\odot$, and follow the cloud through the embedded phase toward gas removal up to $t_{\rm max} \simeq 2.51~t_{\rm ff} \simeq 3.42~\rm Myr$. We define the characteristic free-fall time as
\begin{align}
t_{\rm ff} = \dfrac{\pi R_{\rm GMC}}{2} \sqrt{\dfrac{R_{\rm GMC}}{2 G M_{\rm GMC}}} ,
\label{eq:t_ff}
\end{align}
which differs slightly from that of \cite{Grudic+2021SCFinMCs}. We omit external driving and initialize a uniform magnetic field whose strength is set following \cite{Mouschovias&Spitzer1976BField}:
\begin{align}
\mu_0 = 0.42 \sqrt{\dfrac{|E_G|}{E_B}} ,
\label{eq:B}
\end{align}
where $E_G$ and $E_B$ are the gravitational and magnetic energies of the gas within the cloud, respectively. We adopt $\mu_0 = 4.2$, which corresponds to $\sim 21~\mu \rm G$. Moreover, we initialize $r_{\rm search}$ to $0.01~{\rm pc} \sim 2 \times 10^3~\rm AU$ to better capture wide binaries \citep{Duchene&Kraus2013BinaryStar} and close encounters.

\subsection{Early Spatial Evolution} \label{subsect:evo}

We first examine how collisional dynamics affects the early spatial evolution of the cloud and the star cluster that forms within it. Fig.~\ref{fig:Sigma_T} compares the gas surface density, $\Sigma_{\rm gas}$, and mass-weighted gas temperature, $T_{\rm gas}$, in the \textsc{Arepo-CD} and \textsc{Rigel} runs. The maps show $t = 1.37~\rm Myr \simeq 1.0~t_{\rm ff}$ and $t = 2.06~\rm Myr \simeq 1.5~t_{\rm ff}$, together with gray stellar projections overlaid. At both epochs, the simulations show broadly similar global evolution. Gas filaments feed star formation in dense regions and hot, low-density cavities emerge as stellar feedback becomes important. At $1.0~t_{\rm ff}$, the gas and stellar distributions are nearly indistinguishable between the two runs. By $1.5~t_{\rm ff}$, modest local differences appear in the stellar distribution, while the overall cluster morphology remains similar. These differences are more apparent in the temperature maps. Stellar dynamics redistributes individual feedback sources, and the locations and shapes of heated regions begin to differ. Thus, collisional dynamics does not strongly alter the large-scale cloud evolution at these early times, but introduces local differences in the stellar distribution and the resulting spatial pattern of feedback \citep[see also][]{Bernard+2025PhantomN, Lahen+2025aSPHGAL+KETJU}.

\subsection{Star Formation Efficiency and Star Formation Rate} \label{subsect:SFH}

\begin{figure}[htbp!] \centering
\includegraphics[width=\linewidth]{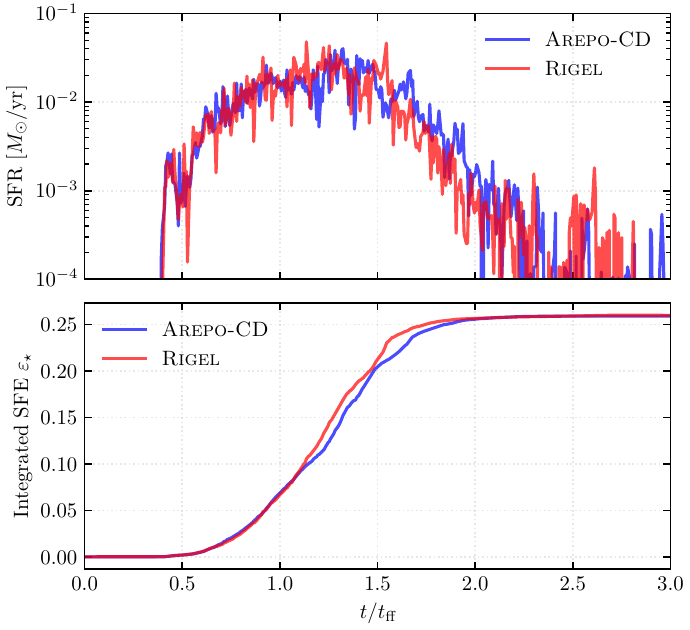}
\caption{SFHs for the star cluster formation simulations in a GMC. Blue curves show the collisional \textsc{Arepo-CD} run and red curves show the collisionless \textsc{Rigel} run. The top panel shows the SFR as a function of time in units of the free-fall time $t_{\rm ff}$ defined in Eq.~\ref{eq:t_ff}. The bottom panel shows the integrated SFE $\varepsilon_\star$ defined in Eq.~\ref{eq:SFE}. The two runs show very similar global evolution, with only modest differences near the peak SFR and during the late-time decline.}
\label{fig:SFH}
\end{figure}

We next quantify the global star formation histories (SFHs) of the two runs through the star formation rate (SFR) and star formation efficiency (SFE). Here we compute the instantaneous SFR from the discrete simulation outputs and divide the stellar mass formed between two successive snapshots by the time interval. The integrated SFE $\varepsilon_\star$, which tracks the evolutionary stage of the cloud, is defined as
\begin{align}
\varepsilon_\star = \dfrac{M_\star}{M_{\rm gas} + M_\star} ,
\label{eq:SFE}
\end{align}
where $M_{\rm gas}$ is the total gas mass of the GMC. Fig.~\ref{fig:SFH} shows the SFHs of the two runs, following a presentation similar to that of \cite{Li+2019}. The top panel shows the same qualitative evolution in both runs. After an initial delay, the SFR rises rapidly, reaches a broad maximum of order $10^{-2}$--$10^{-1}~M_\odot / \rm yr$ around $t \sim 1$--$1.5~t_{\rm ff}$. It then declines as stellar feedback disperses the dense gas. After $t \sim 1.5~t_{\rm ff}$, the SFR decreases steadily to $\lesssim 10^{-3}~M_\odot / {\rm yr}$, with increasing short-timescale variability. The integrated SFE $\varepsilon_\star$ (bottom panel) rises rapidly during the main star-forming phase. It reaches $\sim 0.2$ near the onset of the SFR decline, and then gradually approaches an asymptotic value of $\sim 0.26$. Overall, \textsc{Arepo-CD} and \textsc{Rigel} show very similar SFR and SFE evolution, with only modest differences around the peak SFR and during the late-time decline. This agreement indicates that  large-scale cloud collapse and stellar feedback primarily regulate the global SFH, while the different treatments of stellar dynamics have only a secondary effect on the integrated star formation properties.

\subsection{Velocity Distribution and Runaway Stars} \label{subsect:vel}

\begin{figure}[htbp!] \centering
\includegraphics[width=\linewidth]{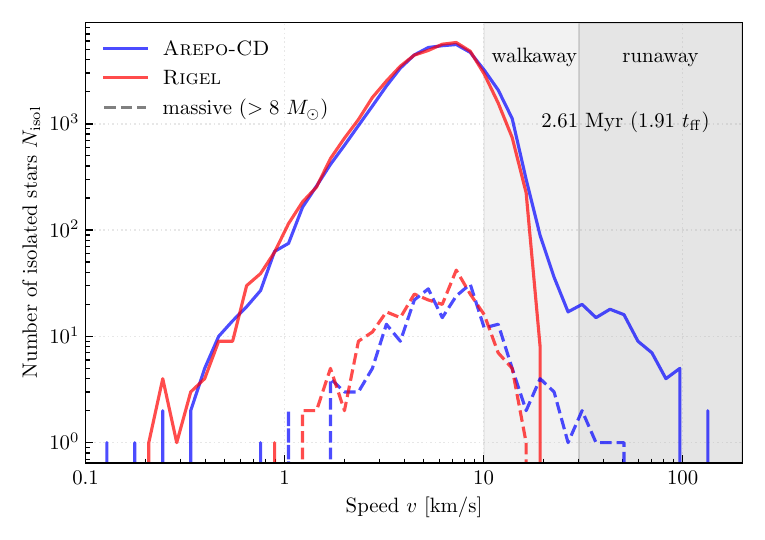}
\caption{Speed distribution in the collisional \textsc{Arepo-CD} run and the collisionless \textsc{Rigel} run at $t = 2.61~{\rm Myr}$ ($1.91~t_{\rm ff}$). We include only isolated stars, defined as stars with no neighbor within $0.01~{\rm pc}$. The solid curves show the full sample of isolated stars, while the ``massive'' curves show the corresponding massive-star subsample ($M_\star > 8~M_\odot$). The bulk distributions are similar, but \textsc{Arepo-CD} develops a much more extended high-velocity tail reaching $\gtrsim 100~{\rm km/s}$. The \textsc{Rigel} distribution declines sharply above a few $\times 10~{\rm km/s}$ and contains far fewer runaway stars.}
\label{fig:vel}
\end{figure}

We next compare the stellar kinematics in the collisional \textsc{Arepo-CD} and collisionless \textsc{Rigel} runs at $t = 2.61~{\rm Myr}$ ($1.91~t_{\rm ff}$). Fig.~\ref{fig:vel} shows the speed distributions for isolated stars and the massive-star subsample ($M_\star > 8~M_\odot$). We include only isolated stars with no neighbor within $0.01~\rm pc$ as in the bottom panels of Fig.~\ref{fig:1k_10k}. The bulk distributions are similar and peak at a few $\rm km/s$, indicating comparable large-scale cluster kinematics. The main difference is the high-velocity tail. \textsc{Arepo-CD} produces numerous fast escapers with speeds reaching $\gtrsim 100~{\rm km/s}$, whereas the \textsc{Rigel} distribution declines sharply above $\sim 10$--$20~{\rm km/s}$. Adopting the conventional threshold of $v \gtrsim 30~{\rm km/s}$ for runaway stars \citep{Eldridge2011runaway&BSS&walkaway, Schoettler+2019runaway&DES&walkaway}, the collisional treatment therefore produces a substantially larger runaway population. This is consistent with close few-body encounters that transfer binding energy into stellar kinetic energy \citep{Spitzer1987}.

\begin{figure*}[htbp!] \centering
\includegraphics[width=\linewidth]{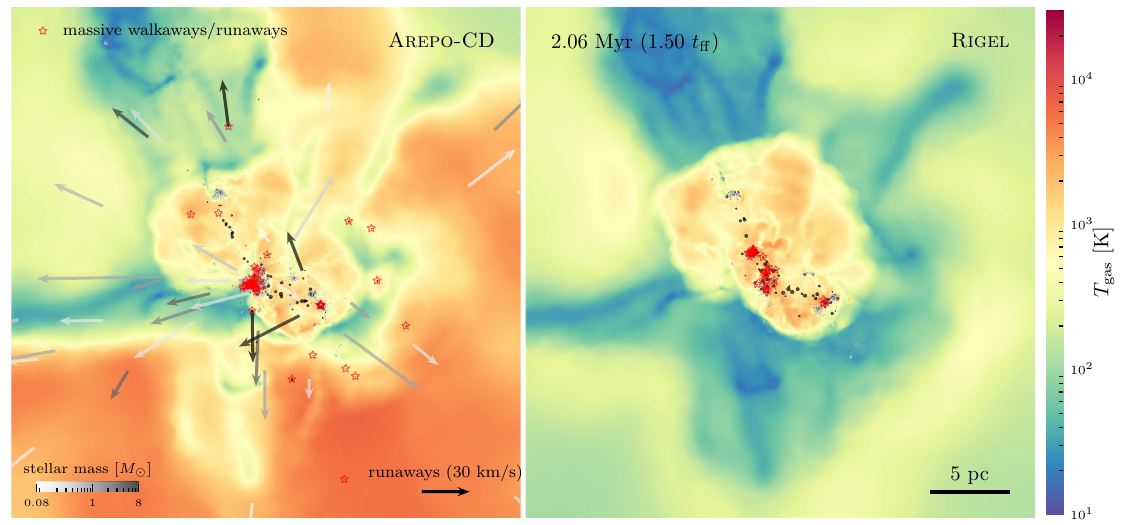}
\caption{Projected gas temperature at $t \simeq 2.06~{\rm Myr}$ ($\simeq 1.5~t_{\rm ff}$) for the collisional \textsc{Arepo-CD} run (left) and the collisionless \textsc{Rigel} run (right). Stellar particles are overlaid and colored by stellar mass. Red star symbols mark massive walkaway and runaway stars ($M_\star \geqslant 8~M_\odot$) with speeds $v > 10~{\rm km/s}$. Velocity vectors show selected high-velocity stars ($v > 30~{\rm km/s}$). We include only isolated stars with no neighbor within $0.01~{\rm pc}$ to exclude high velocities from binary orbital motion. The arrow length indicates the projected stellar speed, with the $30~\rm km/s$ reference vector shown at the lower right of the left panel. Both panels use the same field of view and a $5~\rm pc$ scale bar. Some arrows are shorter than the $30~{\rm km/s}$ reference because stars are selected by their three-dimensional velocities, whereas the plotted vectors show only the projected two-dimensional components. \textsc{Arepo-CD} produces numerous outward-moving escapers through close few-body encounters, while no comparable population is present in \textsc{Rigel}. The ejected massive stars redistribute their feedback, including ionizing radiation and stellar winds. This produces a more extended distribution of heated gas outside the central star-forming region in \textsc{Arepo-CD} and potentially allows more ionizing flux to escape from the embedded region.}
\label{fig:runaway}
\end{figure*}

Fig.~\ref{fig:runaway} shows the spatial distribution of these escapers at $t \simeq 2.06~{\rm Myr}$ ($\simeq 1.5~t_{\rm ff}$). Velocity vectors for stars with $v > 30~{\rm km/s}$ are overlaid on the projected mass-weighted gas temperature, $T_{\rm gas}$. \textsc{Arepo-CD} produces numerous high-velocity stars that move outward from the central star-forming region, whereas \textsc{Rigel} has no comparable population at the same epoch. Because these systems are too young for SN disruption of binaries, these runaways arise from the dynamical ejection processes, including close few-body encounters and subcluster interactions \citep{Poveda+1967runaway&DES, Fujii&PortegiesZwart2011runaway&DES, Farias+2019runaway&DES, Polak+2024bTorch&runaway&DES}.

The massive-star subsamples shown in Figs.~\ref{fig:vel} and \ref{fig:runaway} further demonstrate that the highest velocities are not confined to low-mass stars. This is important because the ejection of massive stars can redistribute stellar feedback. Without such ejections, ionizing sources remain more deeply embedded in the molecular cloud (right panel of Fig.~\ref{fig:runaway}), whereas runaway massive stars can carry ionizing radiation, winds, and later SN feedback into lower-density environments (left panel of Fig.~\ref{fig:runaway}). On galactic scales, this may alter the spatial distribution and escape of ionizing radiation, while displaced massive stars can inject feedback into lower-density gas. There, feedback can couple more efficiently to the ambient ISM \citep{Andersson+2020runaway, Andersson2023INFERNO&runaway}.

\section{Discussion} \label{sect:discuss}

\subsection{Comparison with Existing Hybrid Methods} \label{subsect:compare}

Coupling the collisional $N$-body dynamics of individual stars to hydrodynamic star formation simulations remains technically challenging. Only a few studies address this problem. We compare the performance and results of \textsc{Arepo-CD} with those of existing hybrid frameworks, which generally use one of two approaches. They couple separate codes through an external interface or embed a collisional $N$-body module directly into a hydrodynamic code.

An early example of the first approach is the \textsc{Torch} framework, which couples the grid-based adaptive mesh refinement MHD code \textsc{Flash} to collisional stellar dynamics through \textsc{Amuse} \citep{PortegiesZwart+2009AMUSE, PortegiesZwart2013AMUSE, Pelupessy+2013AMUSE}. Early versions used the fourth-order Hermite integrator \textsc{Ph4} \citep{McMillan+2012ph4, Wall+2019FLASH+ph4, Wall+2020FLASH+ph4}, together with 
the multiple-system module \textsc{Multiples} \citep{PortegiesZwart&McMillan2018AMUSE}
. Later versions replaced \textsc{Ph4} with \textsc{PeTar}. \textsc{Torch} can simulate $\sim 10^6~M_\odot$ clouds and $\gtrsim 10^5$ stars, and has been applied to SFE \citep{Polak+2024aTorch}, runaway stars \citep{Polak+2025Torch}, primordial binaries \citep{Cournoyer-Cloutier+2021PrimordialBinary&Torch, Cournoyer-Cloutier+2024Torch}, and mass segregation \citep{Polak+2025Torch}. Another productive interface-based effort is the SIRIUS (SImulations Resolving IndividUal Stars) project, which developed the tree-direct hybrid SPH/$N$-body code \textsc{Asura+Bridge} \citep{Fujii+2007BRIDGE, Fujii+2021aASURA+BRIDGE, Fujii+2021bASURA+BRIDGE, Fujii+2024aASURA+BRIDGE}. Applications include the formation history of the Orion Nebula Cluster \citep{Fujii+2022aASURA+BRIDGE, Fujii+2022bASURA+BRIDGE}, intermediate-mass black hole formation \citep{Fujii+2024bASURA+BRIDGE}, star cluster assembly \citep{Karam+2025ASURA+BRIDGE, Karam+2026ASURA+BRIDGE}, and primordial binaries \citep{Harada+2026ASURA+BRIDGE} during star cluster formation in molecular clouds. Compared with \textsc{Arepo-CD}, these frameworks can combine existing physics modules more flexibly, for example including modules for primordial binary stars. However, coupling through an external interface can add communication and synchronization overhead. Its effect on efficiency depends on the implementation and the problem. For example, \textsc{Torch} aggregates newly formed stars below a threshold mass to reduce the number of dynamical particles, sacrificing star-by-star sampling at the low-mass end. In contrast, \textsc{Arepo-CD} enables realistic star cluster formation simulations at moderate computational cost. One realization of the fiducial $10^5~M_\odot$ GMC simulation requires only $\sim 3000$ CPU hours,\footnote{A systematic scaling analysis remains to be performed, so this cost does not establish the method's optimal efficiency.} including collisional dynamics and radiative transfer. In terms of scientific outcomes, \textsc{Arepo-CD} can produce similar results for SFE \citep{Polak+2024aTorch}, runaway stars generated by dynamical ejection and subcluster mergers \citep{Polak+2024bTorch&runaway&DES}, and star cluster assembly \citep{Karam+2025ASURA+BRIDGE, Karam+2026ASURA+BRIDGE}.

A complementary strategy is to embed collisional $N$-body methods directly within hydrodynamic codes. \cite{Dinnbier&Walch2020Flash} and \cite{Zimmermann+2025FLASH} implemented a fourth-order Hermite integrator for sink particles in \textsc{Flash}, although without full regularization of few-body systems. Another example is the use of the \textsc{Ketju} module \citep{Rantala2017_KETJU, Mannerkoski2023_KETJU} within \textsc{Sphgal}. \citet{Lahen+2025aSPHGAL+KETJU, Lahen+2025bSPHGAL+KETJU} extended \textsc{Ketju} to resolve collisional dynamics around massive stars with $m_\star > 8~M_\odot$, applying it to idealized dwarf galaxies and to star cluster formation in a low-metallicity dwarf-galaxy merger. Such integrated implementations can reduce the overhead of coupling separate codes, but resolving close encounters only for a selected subset of stars provides an incomplete treatment of collisional dynamics. This approximation may be adequate for galaxy-scale simulations, but it is less suitable for the smaller-scale star cluster simulations considered here. In these systems, interactions among the full stellar population can be dynamically important.

More recent hydrodynamic codes with embedded collisional $N$-body solvers address this limitation by treating collisional dynamics for all stars. One example is \textsc{Enzo}-$N$ \citep{Jo+2024Enzo-Abyss},  which couples the collisional $N$-body code \textsc{Nbody6++gpu} to the cosmological MHD code \textsc{Enzo}. The framework was later rewritten and renamed \textsc{Enzo-Abyss} \citep{Jo+2026Enzo-Abyss, Chung+2026Enzo-Abyss}. This framework targets the formation of (nuclear) star clusters and their coevolution with host galaxies. It also targets high-redshift supermassive black hole seed formation through runaway stellar collisions in dense molecular clouds. 
Similarly, \citet{Bernard+2025PhantomN} implemented a new collisional $N$-body scheme in \textsc{Phantom}. They simulated smaller molecular clouds than those considered here, with a higher gas mass resolution of $\Delta m_{\rm gas} \approx 0.003~M_\odot$. They also emphasized the importance of multiple realizations to reduce the impact of stochastic and chaotic variations in the $N$-body evolution. 
\textsc{Enzo-Abyss} and the \textsc{Phantom} implementation employ a two-level hierarchy combining a fourth-order integrator with a higher-order specialized integrator, whereas \textsc{Arepo-CD} uses a three-level hierarchy that combines second-order leapfrog, fourth-order FSI, and SDAR. This hierarchy applies different integration methods at different dynamical scales, providing greater flexibility for systems spanning a wide dynamical range. The computational cost of the present \textsc{Arepo-CD} simulation also makes multiple realizations and broader parameter surveys practical.

\subsection{Future Extensions} \label{subsect:future}

As a proof-of-concept study, the present work focuses on the main numerical methods and their initial applications. However, the current implementation of \textsc{Arepo-CD} does not yet include all physical processes relevant to young star clusters. We discuss several possible extensions below.

An important future extension of the current framework is the treatment of \textit{stellar collisions and mergers}. In very dense systems, repeated close encounters can lead to direct stellar collisions, whose remnants may evolve very differently from isolated stars \citep{Vergara+2025VMS&IMBH, Vynatheya+2026StarMerger}. A fully self-consistent implementation would require a collision prescription and an evolutionary model for the merger products, including mixing and mass loss. This treatment is computationally demanding, although recent approaches based on machine learning may substantially reduce the cost \citep{AmaroSeoane2026StarMerger&ML, GonzalezPrieto+2026StarMerger&ML, Rose+2026StarMerger&ML}. A more practical first step is to enable the merger support already available in SDAR \citep{Wang+2020SDAR} and then couple it to an analytic prescription for subgrid merger outcomes \citep{Rantala+2026FROST}. This extension would allow \textsc{Arepo-CD} to explore the possibility of runaway stellar collisions and the self-consistent formation of very massive stars (VMSs). These simulations would start from a realistic GMC \citep{Fujii+2024bASURA+BRIDGE} instead of specially constructed pure $N$-body ICs.

A further extension is the inclusion of \textit{primordial binaries}. In the present simulations, compact binaries form only through dynamical interactions, demonstrating that few-body collisional evolution is active. However, real young stellar populations are born with substantial multiplicity. The initial binary fraction, mass-ratio distribution, orbital periods, and eccentricities can strongly affect early cluster heating, merger rates, and runaway production. A realistic primordial-binary module is also likely to increase the collision rate in dense embedded clusters and may promote the formation of VMSs \citep{Rantala+2025FROST}. An important next step is to implement this module consistently with both the IMF and the reservoir-based star formation scheme \citep{Deng+2026aRIGEL}. Encouragingly, pioneering work within the \textsc{Arepo} and \textsc{Rigel} framework has already explored this direction (e.g. G. Liu et al. 2026 in prep.), using mature binary-pairing algorithms \citep{Kouwenhoven+2009PrimordialBinary, Moe&DiStefano2017PrimordialBinary, Cournoyer-Cloutier+2021PrimordialBinary&Torch} that reproduce observed binary trends well \citep{Kroupa1995binary}.

A closely related extension is the inclusion of \textit{binary and multiple stellar evolution (BSE\&MSE)}. The current \textsc{Arepo-CD} implementation follows only single-star evolution through the underlying \textsc{Rigel} framework \citep{Deng+2024RIGEL}. To model interacting binaries, mass transfer, common-envelope evolution, and compact-object binaries formed in dense young clusters, we will ultimately require coupling the subsystem integrator to a BSE or MSE package. This is technically challenging because the relevant dynamical and evolutionary timescales can become comparable in hierarchical few-body systems. Nevertheless, such a coupling will be essential for \textsc{Arepo-CD} to progress from early cluster dynamics to predictive studies of compact-object populations and merger channels. Established options include the classical \textsc{bse} package \citep{Hurley+2002BSE} and more modern alternatives such as \textsc{Sevn} \citep{Spera+2015SEVN} and \textsc{Binary\_c} \citep{Izzard+2018binary_c}. A longer-term option is the secular multiple-evolution code \textsc{mse}, which is explicitly designed to treat arbitrary hierarchical systems and their coupled dynamical and stellar evolution \citep{Hamers+2021MSE}.

Beyond these planned extensions, the framework also points to several immediate scientific applications. The present GMC experiments highlight the roles of feedback timing, runaway production, and early binary formation in shaping embedded clusters. Extending these calculations to systematic surveys over cloud mass, density, metallicity, IMF sampling, and feedback prescriptions will be essential for direct comparison with observed young cluster populations and their scaling relations, such as the mass-radius relation of young star clusters \citep{Brown&Gnedin2021YSC}. Over longer timescales, \textsc{Arepo-CD} should also enable studies of cluster survival, expansion, and tidal disruption in evolving galactic environments, and eventually in cosmological settings. In this sense, this study is a first step toward a predictive framework linking the small-scale physics of close stellar encounters to the large-scale processes of cluster formation and feedback in galaxies.

\section{Summary and Conclusions} \label{sect:summary}

In this paper, we introduced \textsc{Arepo-CD}, a hybrid framework designed to follow star cluster formation in the regime where gas dynamics, stellar feedback, and collisional stellar dynamics operate simultaneously. \textsc{Arepo-CD} couples star-by-star collisional $N$-body dynamics to the moving-mesh MHD code \textsc{Arepo}. It uses unsoftened gravity for star-star interactions while retaining softened gravity for interactions involving gas and other collisionless components. \textsc{Arepo-CD} identifies compact stellar subsystems on-the-fly and evolves them with a hybrid integration scheme combining a fourth-order forward symplectic integrator with slow-down algorithmic regularization. The remaining system follows the native \textsc{Arepo} timestep hierarchy.

We first validated the method by comparing isolated star cluster simulations with those of the collisional $N$-body code \textsc{PeTar}. For the equal-mass model, \textsc{Arepo-CD} reproduces the evolution of the core and Lagrangian radii within the realization-to-realization scatter of \textsc{PeTar}. For clusters with realistic IMFs, the two codes also show close agreement in the radial density profiles and stellar velocity distributions. In particular, \textsc{Arepo-CD} reproduces the high-velocity escaper tail generated by strong collisional interactions, whereas simulations with softened stellar gravity remain more centrally concentrated and produce fewer fast escapers. These tests demonstrate that the hybrid integration scheme captures the main structural and kinematic signatures of collisional cluster evolution.

We then applied \textsc{Arepo-CD} to self-consistent GMC simulations coupled to the \textsc{Rigel} star-formation and feedback model. The collisional and collisionless runs exhibit very similar large-scale gas evolution, star formation rates, and integrated star formation efficiencies, indicating that turbulent collapse and stellar feedback primarily regulate global cloud evolution in the model studied here. Nevertheless, modest local differences develop in the stellar distribution and consequently in the spatial morphology of feedback. A much clearer difference appears in the stellar kinematics. \textsc{Arepo-CD} produces a pronounced high-velocity tail extending beyond $100~{\rm km/s}$ and numerous runaway and walkaway stars, while the collisionless \textsc{Rigel} run produces no comparable runaway population. This demonstrates that strong few-body encounters can already become dynamically important during the embedded phase of cluster formation.

These results establish \textsc{Arepo-CD} as a practical and extensible framework that combines collisional stellar dynamics with hydrodynamic simulations of star cluster formation. Future extensions of the present implementation include stellar collisions and mergers, primordial binaries, and binary stellar evolution. Broader parameter surveys will enable systematic studies of runaway production, feedback redistribution, and the long-term evolution and survival of young star clusters in increasingly complex galactic environments.

\begin{acknowledgments}
We thank Volker Springel for giving us access to \textsc{Arepo}. CK is grateful to Chengzhe Li and Jiani Chu for useful discussions. HL is supported by the National Key R\&D Program of China No. 2023YFB3002502, the National Natural Science Foundation of China under No. 12373006 and 12533004, and the China Manned Space Program with grant No. CMS-CSST-2025-A10. LW acknowledges support from the National Natural Science Foundation of China through grants 12233013 and 12573041, the Fundamental Research Funds for the Central Universities, Sun Yat-sen University (2025QNPY04), the High-level Youth Talent Project (Provincial Financial Allocation) through the grant 2023HYSPT0706. YD is supported by the National Natural Science Foundation of China under No. 125B2057. EMH acknowledges support from the Tsinghua Shui Mu Scholarship and the Beijing Natural Science Foundation (IS25036).
\end{acknowledgments}

\software{\textsc{Arepo} \citep{Springel2010Arepo, Weinberger+2020Arepo}, \texttt{MakeCloud} \citep{Grudic&Guszejnov2021MakeCloud}, \texttt{Matplotlib} \citep{Hunter2007Matplotlib}, \texttt{McLuster} \citep{Kupper+2011McLuster, Wang+2019McLuster}, \texttt{NumPy} \citep{van-der-Walt+2011NumPy}, \textsc{PeTar} \citep{Wang2020PeTar}, \texttt{SDAR} \citep{Wang+2020SDAR}.}

\bibliography{ref}{}
\bibliographystyle{aasjournalv7}



\end{CJK*}
\end{document}